\documentclass[floatfix,%
reprint,
superscriptaddress,
preprintnumbers,
 amsmath,amssymb,
 aps,
prl,
]{revtex4-2}

\usepackage{graphicx}
\usepackage{dcolumn}
\usepackage{braket}
\usepackage[english]{babel}
\usepackage{bm}
\usepackage{comment}
\usepackage{xcolor}
\newcommand{\rev}[1]{{\color{black}#1}}
\newcommand{\revblue}[1]{{\color{black}#1}}
\newcommand{\revorange}[1]{{\color{black}#1}}

\begin{document}

\preprint{APS/123-QED}

\title{How Elasticity and Chiral Damping Shape Domain-Wall Creep}
\author{A. Di Pietro}
\email{a.dipietro@inrim.it}
\affiliation{Istituto Nazionale di Ricerca Metrologica, Torino, Italy}

\author{A. Magni}
\affiliation{Istituto Nazionale di Ricerca Metrologica, Torino, Italy}

\author{S. Pizzini}
\affiliation{Universit{\'e} Grenoble Alpes, CNRS, N{\'e}el Institute,
Grenoble, France}

\author{G. Gandini}
\affiliation{Dipartimento di Fisica, Politecnico di Milano, Via G. Colombo 81, 20131 Milano, Italy }

\author{F. D{\"o}rr}
\affiliation{Department of Physics, Freie Universit{\"a}t Berlin,
Berlin, Germany}

\author{Y. Shokr}
\affiliation{Center for Nuclear Radiation Detectors Application and Research (NÜRDAM), Bolu Abant İzzet Baysal University, Bolu, Türkiye.}
\affiliation{Department of Physics, Freie Universit{\"a}t Berlin,
Berlin, Germany}

\author{M. Erkovan}
\affiliation{Mehmet Tanrıkulu Health Service Vocational School, Bolu Abant İzzet Baysal University, Bolu, Türkiye}
\affiliation{Center for Nuclear Radiation Detectors Application and Research (NÜRDAM), Bolu Abant İzzet Baysal University, Bolu, Türkiye.}

\author{G. Durin}
\affiliation{Istituto Nazionale di Ricerca Metrologica, Torino, Italy}

\author{S. Tacchi}
\affiliation{Istituto Officina dei Materiali del CNR (CNR-IOM),
Sede Secondaria di Perugia, c/o Dipartimento di Fisica e Geologia,
Universit{\`a} di Perugia, Perugia, Italy}

\author{M. Madami}
\affiliation{Dipartimento di Fisica e Geologia,
Universit{\`a} di Perugia, Perugia, Italy}

\author{G. Carlotti}
\affiliation{Dipartimento di Fisica e Geologia,
Universit{\`a} di Perugia, Perugia, Italy}

\author{E. Darwin}
\affiliation{School of Physics and Astronomy,
University of Leeds, Leeds, United Kingdom}
\affiliation{Empa, D{\"u}bendorf, Switzerland}

\author{A. J. Huxtable}
\affiliation{School of Physics and Astronomy,
University of Leeds, Leeds, United Kingdom}

\author{C. Rinaldi}
\affiliation{Dipartimento di Fisica, Politecnico di Milano, Via G. Colombo 81, 20131 Milano, Italy }

\author{C. H. Marrows}
\affiliation{School of Physics and Astronomy,
University of Leeds, Leeds, United Kingdom}

\author{B. J. Hickey}
\affiliation{School of Physics and Astronomy,
University of Leeds, Leeds, United Kingdom}

\author{V.Puliafito}
\affiliation{Department of Electrical and Information Engineering, Politecnico di Bari, 70126, Bari, Italy}

\author{M. Kuepferling}
\affiliation{Istituto Nazionale di Ricerca Metrologica, Torino, Italy}
\date{\today}

\begin{abstract}
Asymmetric domain-wall expansion in the creep regime provides an experimentally accessible probe of the interfacial Dzyaloshinskii-Moriya interaction (iDMI) and, in principle, chiral damping. 
However, severe discrepancies with complementary measurement techniques and the lack of a model describing the full experimental phenomenology of the creep of chiral domain walls, limits the application of these measurements.
Here, using a symmetry decomposition of the velocity curve, we derive an analytical model describing the coupled influence of domain-wall stiffness and chiral damping in the creep regime. Their interplay generates the antisymmetric distortion that shifts the velocity minimum away from the DMI field. Applied to Co-based magnetic multilayers, the model yields iDMI values consistent with BLS and flow-regime benchmarks and identifies the reasons for the longstanding discrepancy. The model provides insights for possible strategies to measure chiral damping via creep dynamics of domain walls.\end{abstract}

\maketitle

The interfacial Dzyaloshinskii-Moriya interaction (iDMI) \cite{MOR-60,DZYALOSHINSKY1958241,fert_role_1980, CRE-98} is a key ingredient in the stabilization and dynamics of chiral magnetic textures in heavy-metal/ferromagnet multilayers with perpendicular magnetic anisotropy, including homochiral domain walls \cite{THI-12,parkin_memory_2015,benitez_magnetic_2015,meng_regulation_2024,di_pietro_domain_2023} and skyrmions \cite{Fert2013,hoffmann_skyrmion-antiskyrmion_2021,mishra_magnetic_2025}. More recently, chiral damping has emerged as a distinct contribution to domain-wall motion, with possible implications for the mobility of magnetic textures \cite{jue_chiral_2016,ganguly_competition_2021,safeer_effect_2022}. The quantitative determination of both effects is therefore relevant to the fundamental understanding of chiral magnetism and to the design of future spintronic devices \cite{das_exploring_2026,zhang_perspective_2026,sankhi_effect_2025,kuepferling_measuring_2023}.

Field-driven domain-wall expansion in the creep regime, in which an out-of-plane field drives the wall in the presence of an in-plane field \cite{Je2013,ganguly_competition_2021,krizakova_study_2019,Lavrijsen2015} holds promise because it couples experimental accessibility and, in principle, sensitivity to both effects \cite{jue_chiral_2016,lau_disentangling_2018}. To date, however, creep measurements have mainly been used to determine the iDMI. In the conventional analysis, the minimum of the domain-wall velocity curve is identified with the DMI effective field, \(H_x^{\min}=-H_{\mathrm{DMI}}\), and the DMI constant is obtained as \(D=\mu_0M_sH_{\mathrm{DMI}}\Delta\), where \(\Delta\) is the domain-wall width parameter \cite{Je2013,hrabec_measuring_2014,kim_wide-range_2017,shahbazi_magnetic_2018}.

\begin{figure}[hbt]
    \centering
    \includegraphics[width=0.9\columnwidth]{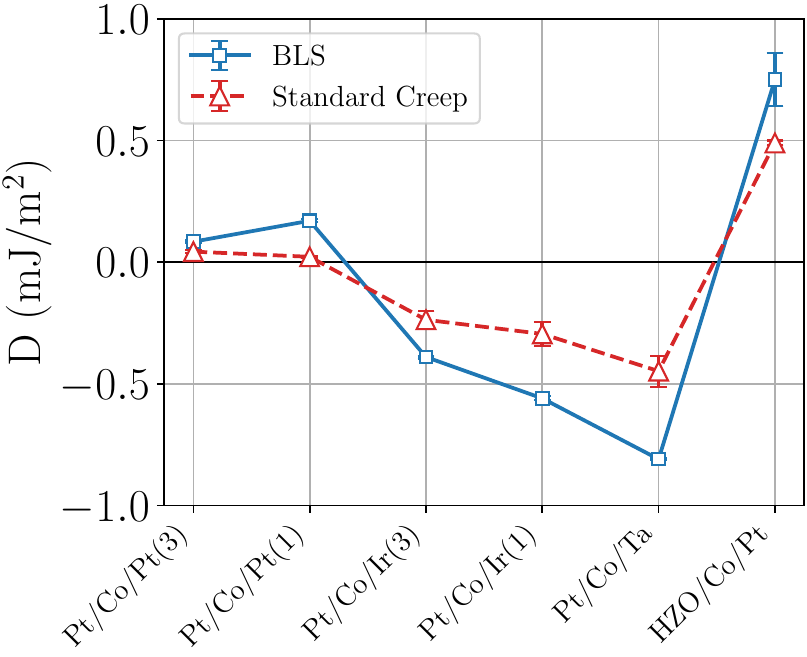}
    \caption{\revblue{iDMI values extracted from BLS and conventional DW creep analysis for the investigated Pt/Co multilayers. The experimental data are from Ref.~\cite{magni_key_2022}; the creep values were obtained using the conventional model of Ref.~\cite{Je2013}. Lines are guides to the eye.}}
    \label{fig:dmi_compare}
\end{figure}

Cross-laboratory comparisons have however highlighted an important discrepancy \cite{magni_key_2022,kuepferling_measuring_2023}: Brillouin light scattering (BLS) determines iDMI through spin-wave nonreciprocity \cite{Cortes-Ortuno2013,YAN-15}, and its estimates agree quantitatively with those obtained from \revblue{DW flow-regime measurements} \cite{garcia_magnetic_2021,vanatka_velocity_2015}; values obtained from \rev{conventional DW creep analysis}, instead, often show substantial over- or underestimates \cite{magni_key_2022,soucaille_probing_2016} (see Fig.~\ref{fig:dmi_compare}). This ambiguity limits both DMI extraction and the use of creep dynamics to probe chiral damping, whose quantitative experimental determination remains challenging. Domain-wall elasticity \cite{pellegren_dispersive_2017,Hartmann2019,DiPietro2026}, and chiral damping \cite{jue_chiral_2016,ganguly_competition_2021} have all been investigated as possible corrections to the conventional creep interpretation, including through numerical treatments \cite{lau_disentangling_2018,davydenko_dzyaloshinskii-moriya_2021}. Nevertheless, an analytical description of the coupled influence of domain-wall elasticity and chiral damping is still missing.

Here, following a symmetry decomposition of the DW velocity curve \cite{kim_decoupling_2025}, we derive a unified model describing the coupled action of domain-wall stiffness and chiral damping. We show why this \rev{discrepancy} is specific to the creep regime and demonstrate that these contributions shift the creep-velocity minimum away from the \rev{DMI field}, \revorange{shedding light on the controversies associated with the conventional $H_{DMI}$ extraction method.}  \cite{Je2013}.

We test this description on five \rev{magnetron-sputtered} \(\mathrm{Pt/Co}/X\) multilayers \revblue{with different capping layers} - Pt/Co/Pt(3), Pt/Co/Pt(1), Pt/Co/Ir(3), Pt/Co/Ir(1), and Pt/Co/Ta - and on a structurally distinct \revblue{\rev{Si/SiO$_2$/TiN(10)/}Hf$_{0.5}$Zr$_{0.5}$O$_2$(10)/Co(0.7)/Pt(6)} sample, denoted HZO/Co/Pt. Growth details, structural and magnetic characterization \rev{of all samples} are reported in the Suppl. Mater. \cite{SupplementalMaterial}. Across the investigated samples, the \rev{symmetry-decomposed DW creep analysis} yields iDMI values consistent with \rev{the BLS benchmarks and, where available, the DW flow benchmarks}. \revorange{The resulting physical picture improves DMI determination from creep measurements and provides insight to extend creep regime measurements to probe chiral damping. }

\begin{figure}
    \centering
    \includegraphics[width=0.6\linewidth]{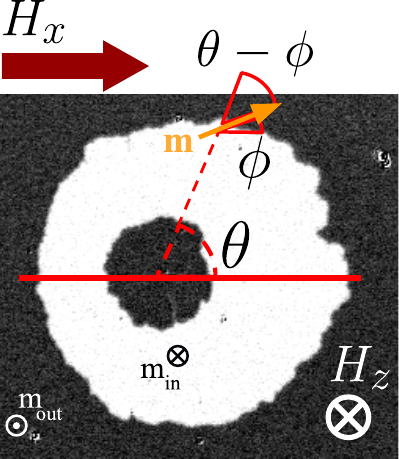}
    \caption{Geometry of asymmetric bubble expansion, defining \(H_x\), \(H_z\), the wall-normal angle \(\theta\), and the internal wall angle \(\phi\).}
    \label{fig:bubble_geom}
\end{figure}

\paragraph{Domain wall creep in the presence of domain wall stiffness and chiral damping}--

We now develop an analytical framework for the combined effects of chiral damping and domain-wall elasticity on creep-regime motion. At low driving fields, the domain-wall velocity is governed by an Arrhenius law \cite{Lemerle1998, chauve_creep_2000,kim_universality_2019},  $v = v_0 \exp[-\chi(H_x)H_z^{-1/4}]$. Here, the dimensionless exponent $\chi(H_x)H_z^{-1/4}$ represents the field-dependent activation energy in units of $k_{\rm B}T$. 
\revorange{In the following, we adopt a semi-phenomenological approximation in which the activation-energy coefficient $\chi(H_x)$ is controlled by the domain-wall stiffness $\tilde{\sigma}$, whereas the prefactor $v_0(H_x)$ depends on the equilibrium wall energy $\sigma_{\rm eq} = \min_\phi \sigma(\theta,\phi)$. Since the stiffness introduces an angular-curvature correction to the equilibrium wall energy, its influence
is amplified in the creep exponent, while its contribution to the algebraic prefactor is expected to be comparatively smaller. We therefore retain this correction in the activation barrier while neglecting it in the prefactor }\cite{Lavrijsen2015}. The domain-wall energy density $\sigma(\theta,\phi)$ is
\begin{align}
    \sigma(\theta,\phi)= &\sigma_0
    -\pi \mu_0 M_s \Delta
    \Bigl[
    H_x \cos\phi
    +
    H_{\rm DMI}\cos(\phi-\theta)
    \Bigr]
    \nonumber\\
    &+
    \frac{\ln (2)}{\pi}t\mu_0 M_s^2
    \cos^2(\phi-\theta).
    \label{eq:DW_density}
\end{align}
where $\theta$ and $\phi$ represent the azimuthal angles of the DW normal and magnetization (see Fig.\ref{fig:bubble_geom}). $\sigma_0 = 4\sqrt{AK_{\mathrm{eff}}}$ is the Bloch wall energy,  $\Delta=\sqrt{A/K_{\mathrm{eff}}}$ is the domain-wall width parameter, $A$ is the exchange stiffness, $K_{\mathrm{eff}} = K_u - \frac{\mu_0 M_s^2}{2}$ is the effective perpendicular magnetic anisotropy, $M_s$ is the saturation magnetization, $H_{\mathrm{DMI}}$ is the DMI field, and $t$ is the effective magnetic thickness. $K_u$ represents the uniaxial magnetocrystalline anisotropy. In this approximation, we write
\begin{equation} 
    \chi(H_x)=\chi_0\left(\frac{\tilde{\sigma}(H_x)}{\tilde{\sigma}(0)}\right)^{1/4},
    \quad 
    v_0 \propto \sigma_{\rm eq}^{\beta}\alpha_{\rm eff}^{-1},
\end{equation}
where $\chi_0$ is the creep scaling coefficient at $H_x=0$, and $\beta$ describes the scaling of the velocity prefactor with the equilibrium domain-wall energy density \cite{akosa_decoupling_2024}.
We model the DW stiffness according to the description of Refs.~\cite{pellegren_dispersive_2017,lau_disentangling_2018},
\begin{equation}
    \tilde{\sigma}(\theta)
    =
    \sigma(\theta,\phi_{\mathrm{eq}})
    +
    \frac{\partial^2\sigma(\theta\,, \phi)}{\partial\theta^2}\bigg|_{\phi=\phi_\mathrm{eq}}.
\end{equation}
Here, $\phi_{\rm eq}= \text{argmin}_{\phi}\sigma(\theta,\phi)$. 
We couple this expression with the chiral damping parametrization \begin{equation}
 \alpha_{\rm eff}(\phi)
 =
 \alpha_0
 \left[
 1+s\eta\cos(\phi-\theta)
 \right],
 \label{eq:CD}
\end{equation}
where \(s=\pm1\) labels the two wall chiralities, \(\eta\) is the dimensionless relative chiral-damping strength, \revblue{and $\alpha_0$ is ordinary Gilbert damping}~\cite{akosa_phenomenology_2016,akosa_decoupling_2024}. Assuming $\theta=0$, the logarithmic velocity, up to irrelevant constants, is
\begin{align}
    \ln v(H_x,H_z)
    &=
    \beta\ln\left(\frac{\sigma_{\rm eq}(H_x)}{\sigma_{\rm eq}(0)}\right)
    -\ln\alpha_{\rm eff}(H_x)
    \nonumber\\
    &\quad
    -\chi_0
    \left(
    \frac{\tilde{\sigma}(H_x)}{\tilde{\sigma}(0)}
    \right)^{1/4}
    H_z^{-1/4}.
    \label{eq:lnv}
\end{align}
To translate this coupled expression into a functional tool, we decompose the velocity curve about the field $-H_{\rm DMI}$, at which the applied in-plane field compensates the DMI contribution to the wall energy \cite{kim_decoupling_2025}. The symmetric component is even and the antisymmetric component is odd about this decomposition center. Under the approximations of small in-plane-field-induced variations of the equilibrium domain-wall energy and stiffness, and retaining chiral damping up to second order $\mathcal{O}(\eta^2)$, Eq.~\eqref{eq:lnv} reduces to a minimal semi-phenomenological model (see \cite{SupplementalMaterial} for the full derivation).
Combining all variables not directly related to $H_{\rm DMI}+H_x$ and defining
$H_K^{DW}=2\ln(2)\,tM_s(\pi^2\Delta)^{-1}$ as the DW anisotropy field, we identify $|H_x+H_{\rm DMI}|\leq H_K^{DW}$ as the mixed Bloch-N\'eel region, in which the equilibrium wall magnetization rotates continuously between the two limiting configurations. The symmetric component of the velocity is governed by two parameters, $\nu_1$ and $\nu_3$:
\begin{widetext}
\begin{align}
&(\ln v)_S =\begin{cases}
    \nu_1(H_K^{DW})^2-\nu_3(H_x+H_{\rm DMI}+H_K^{DW}), 
    & H_x+H_{\rm DMI}<-H_K^{DW},\\[6pt]
    \nu_1(H_x+H_{\rm DMI})^2, 
    & |H_x+H_{\rm DMI}|\le H_K^{DW},\\[6pt]
    \nu_1(H_K^{DW})^2+\nu_3(H_x+H_{\rm DMI}-H_K^{DW}), 
    & H_x+H_{\rm DMI}>H_K^{DW}.
\end{cases}
\label{eq:lnS}
\end{align}
\end{widetext}
The antisymmetric component, under these approximations, is governed by a single parameter $\nu_2$:
\begin{align}
(\ln v)_A =
\begin{cases} 
-\nu_2, 
& H_x+H_{\rm DMI}<-H_K^{DW},\\[6pt]
\nu_2\dfrac{H_x+H_{\rm DMI}}{H_K^{DW}}, 
& |H_x+H_{\rm DMI}|\le H_K^{DW},\\[6pt]
\nu_2, 
& H_x+H_{\rm DMI}>H_K^{DW}.
\end{cases}
\label{eq:lnA}
\end{align}
Introducing the reduced variables
\begin{align}
&\zeta=\frac{\pi\mu_0M_s\Delta H_{\rm DMI}}{\sigma_0},
\quad
\beta_x=\frac{\pi\mu_0M_s\Delta H_x}{\sigma_0}, \\
&\kappa=\frac{\pi \mu_0 M_s \Delta H_K^{DW}}{2\sigma_0},    
\end{align}
and defining $C_z = (\chi_0 H_z^{-1/4})/4$ for compactness, where $C_z$ carries the $H_z$ dependence of the elastic creep contribution, we obtain
\begin{align}
\nu_1
&=
\frac{1}{(H_K^{\rm DW})^2}\bigg[-\frac{\beta\kappa}
{1+\kappa-|\zeta|}
+
\frac{5C_z\kappa}
{1-\kappa}
+
\frac{\eta^2}{2}\bigg],
\label{eq:nu1}
\\
\nu_2
&=
s\eta
-
\frac{\zeta C_z}
{1-\kappa},
\label{eq:nu2}
\\
\nu_3
&=
\frac{2\kappa}{H_K^{\rm DW}}
\left[
\frac{C_z}{1-\kappa}
-
\frac{\beta}{1+\kappa-|\zeta|}
\right].
\label{eq:nu3}
\end{align}
We can now analytically express the shift of the velocity minimum. In the mixed Bloch-Néel region, the velocity minimum $H_x^\star$ is located at
\begin{align}
   (H_x^\star)_{creep}
    &=
    -H_{\rm DMI}
    -
    \frac{\nu_2}{2\nu_1H_K^{DW}}.
    \label{eq:min}
\end{align}
While derived under specific approximations, this expression captures the phenomenology observed experimentally. First, it explicitly shows how the minimum is shifted from the compensation field $-H_{\rm DMI}$ whenever $\nu_2\neq0$. The shift is controlled by the antisymmetric amplitude $\nu_2$, which contains two contributions: 
the chiral-damping term $s\eta$ , and the elastic term $-\zeta C_z/(1-\kappa)$ generated by the DMI-dependent DW stiffness in the activation exponent \revblue{[Eq.~\eqref{eq:nu2}]}. The sign of the chiral-damping term $\propto s\eta$ is sample dependent and not necessarily related to the sign of DMI \cite{safeer_effect_2022,akosa_phenomenology_2016}.
We also highlight the $H_z$ field dependence of the velocity minimum \cite{kim_universality_2019}, which in this formalism can be clearly attributed to the elastic contribution $\sim \zeta C_z$ in $\nu_2$.
\begin{figure}
    \centering
    \includegraphics[width=0.9\linewidth]{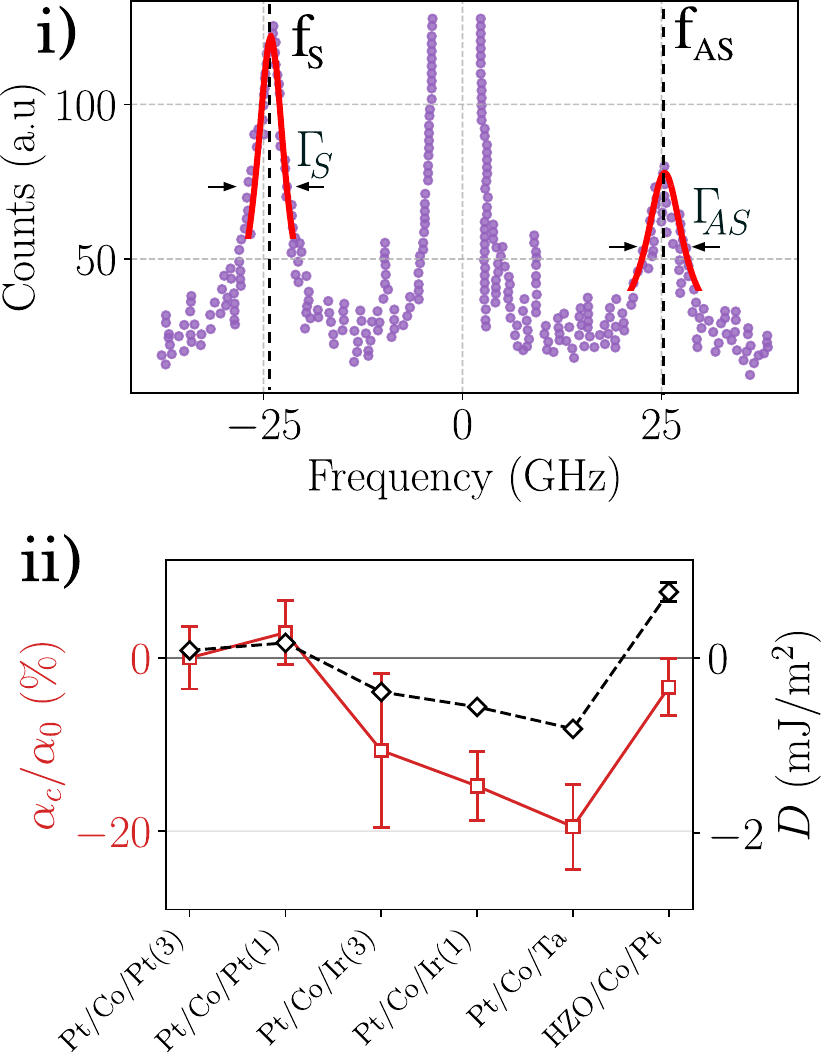}
    \caption{i) Representative BLS spectrum (sample Pt/Co/Ir(1)) showing both the Stokes/anti-Stokes frequencies \(f_{\rm S},f_{\rm AS} \), used to extract \(D\), and the peak-widths  $\Gamma_S,\Gamma_{AS}$. ii) relative chiral damping $\alpha_c/\alpha_0$ contributions obtained from $\Gamma_S,\Gamma_{AS}$ (left axis) and DMI values measured from BLS spectra (right axis). 
}
    \label{fig:2_0}
\end{figure}
\paragraph{Chiral damping and domain-wall elasticity in BLS and \rev{DW flow measurements} -- }
\begin{figure*}[hbt]
    \centering
    \includegraphics[width=0.75\linewidth]{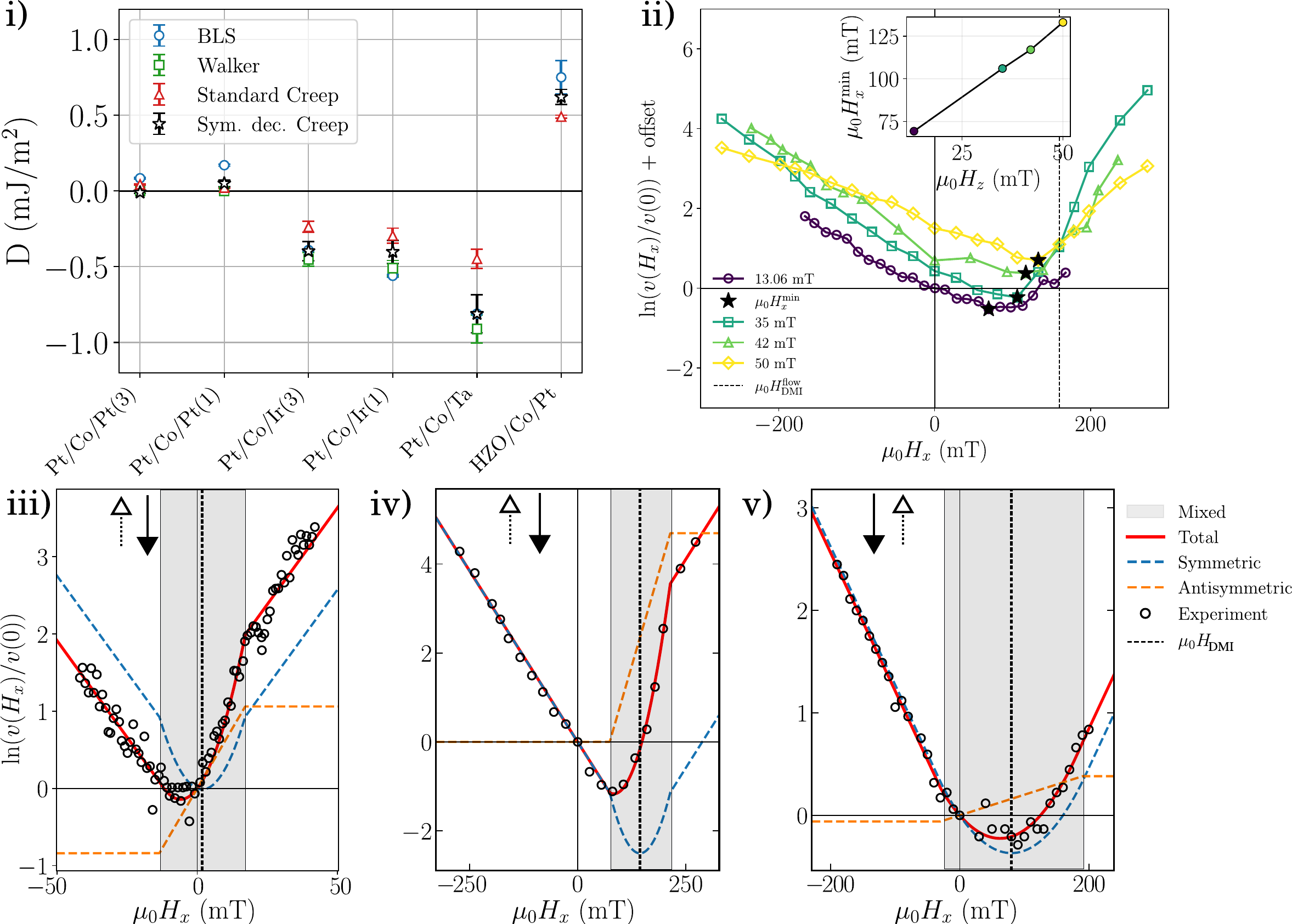}
    \caption{i) iDMI constant values measured with different techniques. No Flow comparison was possible for sample HZO/Co/Pt. ii) Normalized log of the velocity $\ln( v(H_x) / v(0))$ in the asymmetric bubble expansion on sample Pt/Co/Ta performed at different OOP \rev{fields} ranging from $\mu_0H_z = \rev{13}$ mT to $50$ mT. A vertical offset is applied to improve clarity. The inset on the top shows the location of the minimum $\mu_0 H_x^{\mathrm{min}}$ as a function of the applied OOP field $\mu_0 H_z$. Representative velocity \rev{decompositions} according to Eqs.~\eqref{eq:lnS}-\eqref{eq:lnA} for samples iii) Pt/Co/Pt(3), iv) Pt/Co/Ta, and v) HZO/Co/Pt. \rev{The arrows identify the magnetization directions of the two domains and the intervening DW.} The blue dashed line represents the symmetric part \eqref{eq:lnS} of the curve, while the orange dashed line represents its antisymmetric component \eqref{eq:lnA}. The shaded area represents the mixed Bloch-Néel regime.} 
    \label{fig:3} 
\end{figure*}
\label{Sec:Chiral damping and domain wall elasticity in Brillouin light scattering measurements}
\revblue{As shown above, creep domain-wall dynamics are particularly sensitive to the coupled effects of chiral damping and DW stiffness~\cite{akosa_decoupling_2024}. It is therefore important to assess whether the same effects bias BLS and DW flow measurements.}
DW stiffness is not relevant in the interpretation of BLS spectra, as the technique probes propagating spin waves rather than domain-wall motion \cite{nembach_linear_2015,belmeguenai_brillouin_2016}. The \rev{Stokes/anti-Stokes} peak positions and \rev{linewidths} nevertheless contain signatures of both the iDMI and chiral damping. As discussed by Kim \textit{et al.} \cite{kim_chiral_2024}, the iDMI produces a nonreciprocal frequency shift and, through the frequency dependence of ordinary Gilbert damping, also contributes to the linewidth asymmetry\revblue{. Chiral damping, instead,} introduces an additional odd-in-wave-vector contribution to the linewidth asymmetry. \rev{Defining $\Delta f=|f_S|-|f_{AS}|$ where $f_S$ and $f_{AS}$ represent the Stokes and Anti-Stokes peak frequencies [Fig.~\ref{fig:2_0}(i)], the DMI constant is obtained as $D=\pi M_s\Delta f/(2\gamma k)$, where $k$ is the spin-wave wave vector and $\gamma$ is the gyromagnetic ratio in frequency units, for the sign convention used here.} Defining the measured linewidth sum and difference $\Gamma_{\pm}=\Gamma_S \pm \Gamma_{AS}$ [Fig.~\ref{fig:2_0}(i)]\rev{,} we can isolate the chiral-damping contribution via
\begin{equation}
    \frac{\alpha_c}{\alpha_0}
    =
    \frac{\Gamma_--r_f\Gamma_+}
    {\Gamma_+-r_f\Gamma_-},
    \label{eq:Gamma_S_AS}
\end{equation}
where $r_f=(|f_S|-|f_{AS}|)/(|f_S|+|f_{AS}|)$ \cite{SupplementalMaterial}. \revblue{Here, $\alpha_c$ denotes the chiral contribution to the damping.}
\rev{The resulting chiral-damping contributions for our sample series are shown in Fig.~\ref{fig:2_0}(ii).} Within the \(\mathrm{Pt/Co}/X\) series, the sign and magnitude of the linewidth asymmetry broadly track those of $D$. This \revblue{suggests} a sample-dependent correlation rather than a general proportionality: DMI and chiral damping correspond to distinct energetic and dissipative responses and can vary independently~\cite{kim_chiral_2024}, as illustrated here by the weak linewidth asymmetry of sample HZO/Co/Pt despite its sizable DMI.

\paragraph{\rev{DW flow measurements} --}
The domain-wall stiffness correction governing the creep activation barrier does not enter the flow regime, where the motion is no longer controlled by thermally activated depinning \cite{metaxas_creep_2007,slonczewski_dynamics_1972}. In the strong-DMI limit, the Walker-velocity plateau provides a direct estimate of the DMI strength \cite{ha_pham_very_2016},
\begin{equation}
v_W \simeq \frac{\pi}{2}\gamma\frac{D}{M_s},
\label{eq:vW}
\end{equation}
Since Eq.~\eqref{eq:vW} contains no damping parameter at leading order, this estimate is insensitive to chiral damping within the present approximation \cite{akosa_phenomenology_2016,akosa_decoupling_2024}.
A second \rev{DW flow estimate} of \(H_{\rm DMI}\) is obtained from the minimum of the precessional-flow velocity as a function of \(H_x\) \cite{garcia_magnetic_2021}. \rev{Chiral damping weakly breaks the symmetry of this velocity curve.} Expanding the average drift velocity \cite{garcia_magnetic_2021,vanatka_velocity_2015} to first order in \(\eta\) yields the following $(H_x^\star)_{\mathrm{flow}}$ field corresponding to the velocity minimum
\begin{equation}
(H_x^\star)_{\mathrm{flow}}
\simeq -H_{\mathrm{DMI}}
+\frac{s\eta\alpha_0H_z}{\pi}
\left[
1+\frac{5}{8}
\left(\frac{H_K^{DW}}{H_z}\right)^2
\right].
\label{eq:flow_shift}
\end{equation}
The derivation and comparison with the full numerical minimum are given in the Supplementary material \cite{SupplementalMaterial}. Equation~\eqref{eq:flow_shift} shows that the displacement is suppressed by \(\eta\alpha_0\). For the parameters of the investigated samples, the calculated shift is of the order of \(1\)~mT and is comparable to the experimental resolution of the in-plane-field coils. To summarize, \rev{domain-wall} elasticity does not enter the flow physics, the Walker velocity is damping independent to leading order, and the chiral-damping-induced displacement of the drift-velocity minimum is experimentally negligible. We conclude that \rev{DW flow measurements} are expected to agree with BLS, providing a reference for the present analysis of iDMI strength \cite{ha_pham_very_2016}.

\paragraph{Experimental interpretation --}
\revblue{We now apply the symmetry-decomposed DW creep model of Eqs.~\eqref{eq:lnS}-\eqref{eq:lnA} to the samples introduced above. We test whether the model yields iDMI values consistent with BLS and DW flow measurements and whether the fitted antisymmetric response helps identify the physical origin of the discrepancies found with conventional DW creep analysis.}

\revblue{The symmetry-decomposed estimates are obtained by fitting each measured $(\ln[v(H_x)/v(0)])$ as
\begin{equation}
\ln v(H_x)=(\ln v)_S+(\ln v)_A,
\label{eq:full_response}
\end{equation}
where $(\ln v)_S,(\ln v)_A$ come from Eqs.\eqref{eq:lnS}-\eqref{eq:lnA}. We see how \(H_x=-H_{\rm DMI}\) is \rev{the symmetry axis} of $(\ln v)_S$, while $(\ln v)_A$ displaces the velocity minimum from this center. In Fig.~\ref{fig:3}(i), The blue circles represent BLS measurements, the green squares are the \rev{DW flow estimate} extracted from the Walker-velocity plateau using Eq.~\eqref{eq:vW}, the red triangles are the \rev{conventional DW creep estimate}~\cite{Je2013}, and the black stars are the \rev{symmetry-decomposed DW creep estimate} [Eq.~\eqref{eq:full_response}].  We immediately notice how the symmetry-decomposed DMI estimates agree with independent BLS measurements and, where available, DW flow estimates. The conventional DW creep estimates, on the other hand, show substantial discrepancies.} \rev{An in depth discussion of the error estimation of $D = \mu_0 M_s \Delta H_{\mathrm{DMI}}$ for each sample as well as the fitted values of \(\nu_{1,2,3}\) and \(H_K^{DW},H_{\mathrm{DMI}}\) are reported in \cite{SupplementalMaterial}.} We can now discuss some representative cases: sample \(\mathrm{Pt/Co/Pt}(3)\) [Fig.~\ref{fig:3}(iii)] has a small antisymmetric component and a symmetric part centered close to zero, with a weakly biased \rev{conventional DW creep estimate}. Sample \(\mathrm{Pt/Co/Ta}\) [Fig.~\ref{fig:3}(iv)] has a large antisymmetric component and a symmetric part centered at large \(H_x\), with a large displacement of the minimum from the symmetry axis. Sample HZO/Co/Pt [Fig.~\ref{fig:3}(v)] combines a small antisymmetric component with a symmetric part centered at large \(H_x\), yet \revblue{retains a finite discrepancy between the conventional DW creep and BLS estimates [inset of Fig.~\ref{fig:3}(v)]}. The inverted sign of $H_{\mathrm{DMI}}$ in this sample follows from the inverted layer stacking: Pt is on top of Co, in contrast to the \(\mathrm{Pt/Co}/X\) series. 

Comparing these decompositions with the BLS spectra provides further insights. As shown in Fig.~\ref{fig:2_0}(i), the spectra provide both the Stokes/anti-Stokes frequency shift used to extract DMI and the relative linewidth asymmetry \(\Gamma_-\), \rev{which we associate with chiral damping}~\cite{kim_chiral_2024,SupplementalMaterial} via Eq.\eqref{eq:Gamma_S_AS}. Across the \(\mathrm{Pt/Co}/X\) series [Fig.~\ref{fig:2_0}(ii)], the largest linewidth asymmetries occur in the samples where the discrepancy between \rev{conventional DW creep analysis} and BLS is largest. However, sample HZO/Co/Pt in Fig.~\ref{fig:3}(v) shows that this correspondence is not one-to-one: a sizable minimum displacement persists at large DMI even when the BLS chiral damping signature is weak [Fig.~\ref{fig:2_0}(ii)], consistent with an elastic contribution to the asymmetry $\nu_2$. Its discrepancy is nevertheless smaller than that of \(\mathrm{Pt/Co/Ta}\), but accounting for the asymmetric part yields a DMI value closer to the BLS benchmark. \rev{Combining these observations, we conclude that elastic and chiral damping contributions mix with sample-dependent weights, obscuring the origin of the discrepancy in the full creep velocity curve.}

\revblue{We emphasize, in fact,  how the magnitude of the antisymmetric component alone cannot establish the relative roles of DW elasticity and chiral damping.} At a single \(H_z\), the fit determines the net \(\nu_2\) \rev{of Eq.~\eqref{eq:nu2}}, but does not separate its elastic and dissipative contributions. Such a separation would require extracting \(\nu_2\) from velocity curves measured over a range of \(H_z\), since the elastic term varies through \(C_z\propto H_z^{-1/4}\).

The different \(H_z\) dependencies of elasticity and chiral damping in $\nu_2$ (see Eq.~\eqref{eq:nu2}) provide a way to test this interpretation. Figure~\ref{fig:3}(ii) shows that the conventional DW creep minimum is not a fixed material observable. In sample \(\mathrm{Pt/Co/Ta}\), the normalized DW creep curves \(\ln[v(H_x)/v(0)]\), measured over OOP fields spanning \(\mu_0H_z=\rev{13}\)-\(50\) mT (all values below the depinning field for sample \(\mathrm{Pt/Co/Ta}\), $\mu_0 H_d = 55.3 \pm 0.3$ mT \cite{SupplementalMaterial}), display a systematic displacement of \(\mu_0H_x^{\rm min}\) with the out-of-plane driving field. This trend, summarized in the inset, is consistent with Eq.~\eqref{eq:min}: changing \(H_z\) modifies the activation-barrier contribution through \(C_z\), and therefore changes the relative weight of the elastic term in the shift of the velocity minimum. 
\paragraph{Conclusion - }
In this work, we analyzed how domain-wall elasticity and chiral damping collectively affect the creep dynamics of chiral domain walls. The derived model shows that the antisymmetric component responsible for shifting the \rev{domain-wall creep-velocity minimum away from the DMI field} arises from a nontrivial interplay of these contributions. Moreover, because the domain-wall stiffness contribution depends on the out-of-plane driving field, the position of \rev{this minimum} can vary with the measurement conditions rather than representing an intrinsic material parameter. Fits to velocity curves measured in a series of Pt/Co samples with different overlayers and growth procedures yield DMI values consistent with \rev{BLS and the available DW flow benchmarks}, whose iDMI signatures are not affected by the same interplay between elasticity and chiral damping. \revorange{With additional measurement of the $H_z$ dependence of the $\nu_2$ component of the velocity curves (see. Eq.\eqref{eq:nu2}), creep measurements could also serve as a probe of chiral damping in magnetic thin films, providing a basis for future quantitative comparisons with BLS.}
\paragraph{Acknowledgments --}
We acknowledge R. Lavrijsen for the fruitful discussions.  This project is supported by the European Union (EU) and the Italian ministry of University and Research (MUR) under the grant PRIN 2022 “Metrology for spintronics: A machine learning approach for the reliable determination of the Dzyaloshinskii-Moriya interaction (MetroSpin)”. Project code: 2022SAYARY. Yasser Shokr gratefully acknowledges financial support from TÜBİTAK under
grant number 124C509, which supported his work. This work was supported by the France 2030 government
plan managed by the Agence Nationale de la Recherche (project PEPR
SPIN CHIREX ANR‐22‐EXSP‐0002). This work was supported in part by the European Metrology Programme for Innovation and Research (EMPIR) Programme co-financed by the Participating States under Project 17FUN08-TOPS.

\end{document}


\title{Supplementary Material: How Elasticity and Chiral Damping Shape Domain-Wall Creep}

\author{A. Di Pietro}
\email{a.dipietro@inrim.it}
\affiliation{Istituto Nazionale di Ricerca Metrologica, Torino, Italy}

\author{A. Magni}
\affiliation{Istituto Nazionale di Ricerca Metrologica, Torino, Italy}

\author{S. Pizzini}
\affiliation{Universit{\'e} Grenoble Alpes, CNRS, N{\'e}el Institute,
Grenoble, France}

\author{G. Gandini}
\affiliation{Dipartimento di Fisica, Politecnico di Milano, Via G. Colombo 81, 20131 Milano, Italy }

\author{F. D{\"o}rr}
\affiliation{Department of Physics, Freie Universit{\"a}t Berlin,
Berlin, Germany}

\author{Y. Shokr}
\affiliation{Center for Nuclear Radiation Detectors Application and Research (NÜRDAM), Bolu Abant İzzet Baysal University, Bolu, Türkiye.}
\affiliation{Department of Physics, Freie Universit{\"a}t Berlin,
Berlin, Germany}

\author{M. Erkovan}
\affiliation{Mehmet Tanrıkulu Health Service Vocational School, Bolu Abant İzzet Baysal University, Bolu, Türkiye}
\affiliation{Center for Nuclear Radiation Detectors Application and Research (NÜRDAM), Bolu Abant İzzet Baysal University, Bolu, Türkiye.}

\author{G. Durin}
\affiliation{Istituto Nazionale di Ricerca Metrologica, Torino, Italy}

\author{S. Tacchi}
\affiliation{Istituto Officina dei Materiali del CNR (CNR-IOM),
Sede Secondaria di Perugia, c/o Dipartimento di Fisica e Geologia,
Universit{\`a} di Perugia, Perugia, Italy}

\author{M. Madami}
\affiliation{Dipartimento di Fisica e Geologia,
Universit{\`a} di Perugia, Perugia, Italy}

\author{G. Carlotti}
\affiliation{Dipartimento di Fisica e Geologia,
Universit{\`a} di Perugia, Perugia, Italy}

\author{E. Darwin}
\affiliation{School of Physics and Astronomy,
University of Leeds, Leeds, United Kingdom}
\affiliation{Empa, D{\"u}bendorf, Switzerland}

\author{A. J. Huxtable}
\affiliation{School of Physics and Astronomy,
University of Leeds, Leeds, United Kingdom}

\author{C. Rinaldi}
\affiliation{Dipartimento di Fisica, Politecnico di Milano, Via G. Colombo 81, 20131 Milano, Italy }

\author{C. H. Marrows}
\affiliation{School of Physics and Astronomy,
University of Leeds, Leeds, United Kingdom}

\author{B. J. Hickey}
\affiliation{School of Physics and Astronomy,
University of Leeds, Leeds, United Kingdom}

\author{V.Puliafito}
\affiliation{Department of Electrical and Information Engineering, Politecnico di Bari, 70126, Bari, Italy}

\author{M. Kuepferling}
\affiliation{Istituto Nazionale di Ricerca Metrologica, Torino, Italy}
\date{\today}

\maketitle

\section{Sample preparation and magnetic characterization}

\subsection{Previously characterized Co-based samples}
\label{sec:samples_a1_a5}
\noindent Samples a1-a5 are ultrathin Co-based heterostructures previously
prepared and characterized in Ref.~\cite{magni_key_2022}. All samples
contain a nominally $0.8$-nm-thick Co layer deposited on a common
Ta(5)/Pt(3) underlayer, while the material and thickness of the upper
interface were varied. The nominal layer sequences, listed from bottom
to top and with all thicknesses expressed in nanometers, are reported in Table.~\ref{tab:samples_Leeds}
The complete fabrication and initial characterization were reported previously in
Ref.~\cite{magni_key_2022}; here we summarize only the information
required for the analysis presented in this work.

\subsection{HZO sample}
The a6 sample consists of an SiO/TiN/Hf$_{0.5}$Zr$_{0.5}$O$_2$(10)/Co(0.7)/Pt(6.0) stack fabricated at the Fraunhofer Institute Dresden and characterized at the Politecnico di Milano. Atomic Layer Deposition (ALD) was used for TiN and Hf$_{0.5}$Zr$_{0.5}$O$_2$ (HZO), while in-situ Physical Vapour Deposition (PVD) was used for the metallic layers (Co/Pt).

\begin{table}[t] \centering \caption{Layer structures of the investigated samples. Layer thicknesses are given in nanometers.} \label{tab:layer_structures} \begin{tabular}{cl} \hline \hline Sample ID & Layer structure \\ \hline a1 & $\mathrm{Ta}(5)/\mathrm{Pt}(3)/\mathrm{Co}(0.8)/ \mathrm{Pt}(3)/\mathrm{Ta}(3)$ \\ a2 & $\mathrm{Ta}(5)/\mathrm{Pt}(3)/\mathrm{Co}(0.8)/ \mathrm{Pt}(1)/\mathrm{Ta}(3)$ \\ a3 & $\mathrm{Ta}(5)/\mathrm{Pt}(3)/\mathrm{Co}(0.8)/ \mathrm{Ir}(3)/\mathrm{Ta}(3)$ \\ a4 & $\mathrm{Ta}(5)/\mathrm{Pt}(3)/\mathrm{Co}(0.8)/ \mathrm{Ir}(1)/\mathrm{Ta}(3)$ \\ a5 & $\mathrm{Ta}(5)/\mathrm{Pt}(3)/\mathrm{Co}(0.8)/ \mathrm{Ta}(3)$ \\ a6 & Hf$_{0.5}$Zr$_{0.5}$O$_2$(10)/Co(0.7)/Pt(6.0) \\ \hline \hline \label{tab:samples_Leeds} \end{tabular} \end{table}

\begin{table}
\begin{threeparttable}[b]
\caption{Saturation magnetization $M_s$, Exchange stiffness $A$, Domain wall (DW) width parameter $\Delta$, effective anisotropy $K_{\mathrm{eff}}$  for the full sample series of Tab.\ref{tab:layer_structures}. }
\begin{tabular}{lcccc}
\hline
\hline
Sample ID
  & $M_s$& $A$
  & $\Delta$& $K_{\mathrm{eff}}$\\
  & (MA/m) & (pJ/m) & (nm) & (MJ/m$^3$) \\
\midrule
a1 & $1.58 \pm 0.03^{b}$& $13 \pm 2^b$& $4.8 \pm 0.4^b$& $0.490\pm 0.02^b$\\
a2 & $1.74 \pm 0.04^{b}$& $14 \pm 2^b$& $5.5 \pm 0.8^b$& $0.420 \pm 0.03^b$\\
a3 & $1.14 \pm 0.05^{b}$& $13\pm4^a$& $5.6 \pm 0.8^a$& $0.414 \pm 0.001^b$\\
a4 & $1.20 \pm 0.09^{b}$&  $12\pm4^a$& $4.8 \pm 0.8^a$& $0.513 \pm 0.002^b$\\
a5 & $0.90 \pm 0.02^{b}$& $23\pm6^a$& $6.3 \pm 0.9^a$& $0.563 \pm 0.005^b$\\

 a6& $1.42 \pm 0.02$ & $20 \pm 5^{c}$& $5.4 \pm 0.9 $ & $0.69 \pm 0.07$ \\
 \hline
\hline
 \end{tabular}
\begin{tablenotes}
\item [a] Obtained via measurements in the flow regime as discusses in Sec.\ref{sec:flow_characterization} \cite{ha_pham_very_2016}
\item [b] Obtained from the measurements in Ref.~\cite{magni_key_2022}
\item [b] Literature value on similar composition stack~\cite{Zimmermann2019,devolder_exchange_2016}
\end{tablenotes}
\end{threeparttable}

\end{table}

\section{Brillouin light scattering measurements}
\label{sec:BLS}

\subsection{DMI evaluation}

\noindent Independent measurements of the interfacial Dzyaloshinskii-Moriya
interaction were performed by Brillouin light scattering on samples
taken from the same wafers as those used for the domain-wall
measurements.

\begin{figure}
    \centering
    \includegraphics[width=0.8\linewidth]{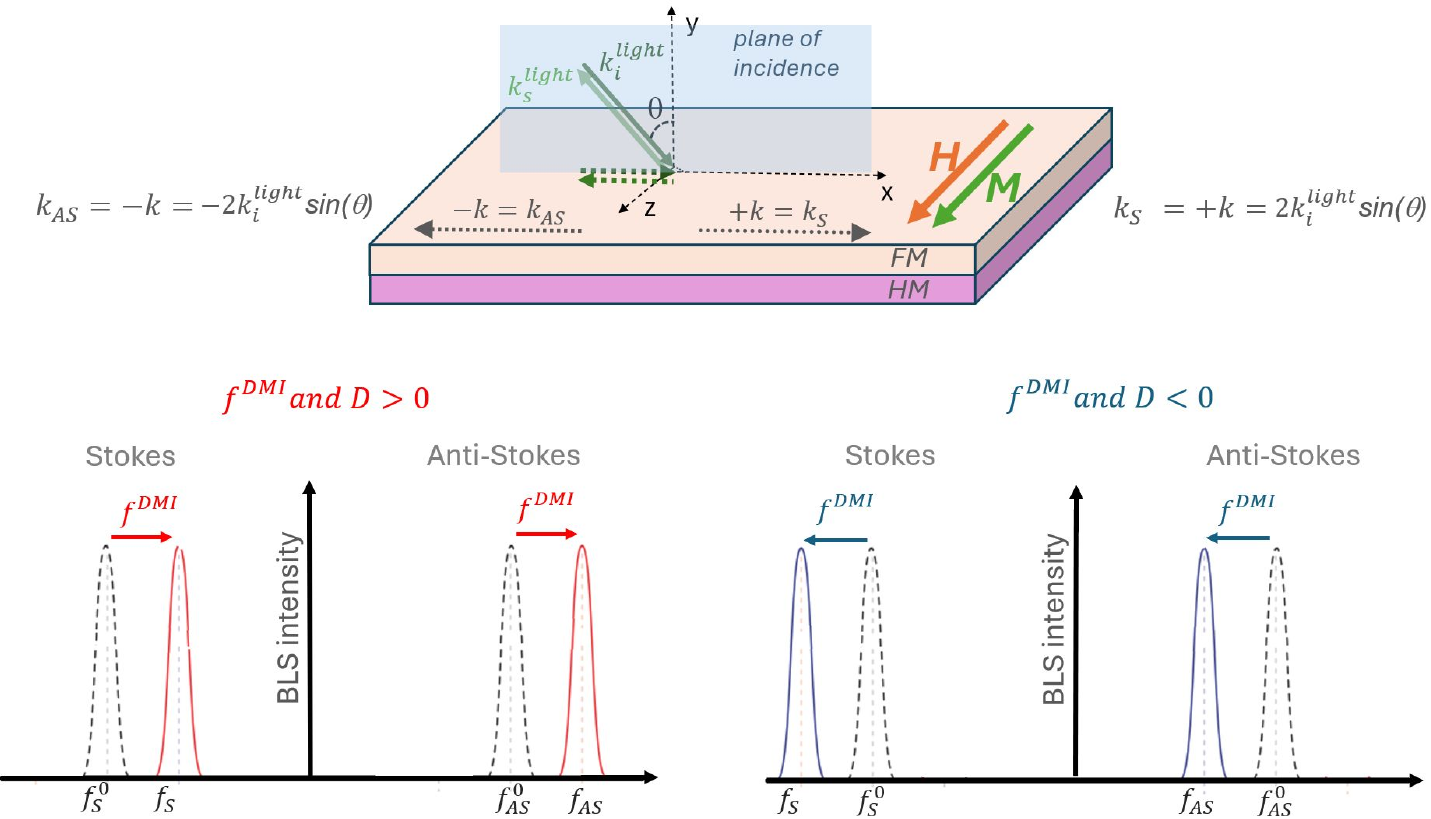}
    \caption{Schematic diagram of the BLS interaction setup. The bottom panel shows a typical BLS spectrum in the absence of iDMI (dashed line) and with iDMI (solid line) for negative (left) and positive (right) values of $D$, respectively. Here $k_{\mathrm{i}}^{\mathrm{light}}$ denotes the incident light wave numbers, while $H$ and $M$ are the applied in-plane field and magnetization, respectively.}
    \label{fig:BLS_Scheme}
\end{figure}

In a typical BLS experiment, a monochromatic laser beam is focused onto the sample surface, and the backscattered light is collected by the same objective and frequency analyzed by a high-resolution spectrometer. BLS measurements were performed in the Damon–Eshbach (DE) geometry by applying an external magnetic field $H$, sufficiently large to magnetize the ferromagnetic film in-plane, and sweeping the spin-wave wave vector $k$ perpendicular to the magnetization direction (Fig.~\ref{fig:BLS_Scheme}). During the BLS scattering process, magnons can be either created (Stokes process) or annihilated (anti-Stokes process), resulting in a redshift or blueshift of the scattered light relative to the incident beam. In wave-vector space, Stokes and anti-Stokes processes correspond to magnons with opposite wave vectors. In the presence of DMI, the frequency difference between DE modes propagating in opposite direction can be written as:

\begin{equation}
\Delta f(k)=f_S-f_{AS}
          = \frac{\gamma_0 D}{\pi\ M_s}k,
\label{eq:BLS_DMI}
\end{equation}

where $k$ is the in-plane spin-wave wave vector and $\gamma_0$ is the gyromagnetic ratio. The sign of $D$ was assigned using the field,
wave-vector, and Stokes/anti-Stokes conventions described below.
The BLS measurements probe an area with a lateral dimension of
approximately $10$-$40~\mu$m and spin waves with wavelengths of the
order of $0.5~\mu$m. The technique therefore provides a spatially
averaged measurement that is complementary to domain-wall creep,
which is particularly sensitive to disorder and pinning on nanometer
length scales.

\subsection{Extraction of chiral damping from the BLS linewidths}

\noindent The Stokes and anti-Stokes components of each Brillouin-light-scattering spectrum were fitted independently with Lorentzian line shapes. Their linewidths, denoted by $\Gamma_{\mathrm{S}}$ and $\Gamma_{\mathrm{AS}}$, characterize the dissipation of counterpropagating spin waves. Following Kim \textit{et al.}~\cite{kim_chiral_2024}, the linewidths for the field, magnetization, and wave-vector convention adopted here can be written as

\begin{equation}
\Gamma_{\mathrm{S}}=C(\alpha_0+\alpha_c)(1+r_f),
\qquad
\Gamma_{\mathrm{AS}}=C(\alpha_0-\alpha_c)(1-r_f),
\label{eq:BLS_linewidth_model}
\end{equation}
where $C$ is the common wave-vector-even linewidth prefactor, $\alpha_0$ is the ordinary Gilbert-damping coefficient, and $\alpha_c$ is its wave-vector-odd contribution. The DMI frequency contribution is determined from

\begin{equation}
r_f=
\frac{|f_{\mathrm{S}}|-|f_{\mathrm{AS}}|}
{|f_{\mathrm{S}}|+|f_{\mathrm{AS}}|}
=\frac{\omega_{\mathrm{DMI}}}{\omega_0}.
\label{eq:BLS_frequency_asymmetry}
\end{equation}
We define the sum and difference of the measured linewidths as

\begin{equation}
\Gamma_{+}=\Gamma_{\mathrm{S}}+\Gamma_{\mathrm{AS}},
\qquad
\Gamma_{-}=\Gamma_{\mathrm{S}}-\Gamma_{\mathrm{AS}}.
\end{equation}
Equation~\eqref{eq:BLS_linewidth_model} then gives

\begin{equation}
\Gamma_{+}=2C(\alpha_0+r_f\alpha_c),
\qquad
\Gamma_{-}=2C(\alpha_c+r_f\alpha_0).
\end{equation}
Eliminating the common prefactor $C$ yields the relative chiral-damping contribution directly:

\begin{equation}
\frac{\alpha_c}{\alpha_0}
=
\frac{\Gamma_{-}-r_f\Gamma_{+}}
{\Gamma_{+}-r_f\Gamma_{-}}
.
\label{eq:BLS_relative_chiral_damping}
\end{equation}
This expression removes the DMI-induced contribution.

\begin{table*}[t]
    \centering
    \caption{Stokes ($\Gamma_S$) and anti-Stokes linewidths ($\Gamma_{AS}$), fitted DMI frequency splitting $\Delta f_{\mathrm{DMI}}$, normalized frequency asymmetry $r_f$ and extracted relative chiral damping $\alpha_c/\alpha_0$. }
    \label{tab:BLS_chiral_damping}
    \begin{tabular}{lccccc}
        \hline
        \hline
        Sample & $\Gamma_{\mathrm{S}}$ & $\Gamma_{\mathrm{AS}}$ & $\Delta f_{\mathrm{DMI}}$ & $r_f$ & $\alpha_c/\alpha_0$ \\
         & (GHz) & (GHz) & (GHz) & (-) & (\%)
        \\
        \hline
a1 & $3.0 \pm 0.1$& $3.0 \pm 0.2$& $0.1 \pm 0.3$& $0.4\pm 0.9$& $0 \pm 4$\\
a2 & $3.2 \pm 0.1$& $3.0 \pm 0.2$& $0.2 \pm 0.2$& $0.4 \pm 0.4$& $3 \pm 4$\\
a3 & $4.3 \pm 0.3$& $5.5 \pm 0.9$& $-0.7 \pm 0.5$& $-2 \pm 1$& $-11\pm 9$\\
a4 & $5.0\pm 0.2$& $7.0 \pm 0.5$& $-0.9 \pm 0.3$& $-1.9 \pm 0.6$& $-15 \pm 4$\\
a5 & $4.6 \pm 0.2$& $7.3 \pm 0.7$& $-1.9 \pm 0.2$& $-3.8 \pm 0.5$& $-20 \pm 5$\\
a6 & $12.1 \pm 0.5$& $13.7 \pm 0.7$& $1.1 \pm 0.1$& $2.8\pm 0.3$& $-3 \pm 3$\\
        \hline\hline
    \end{tabular}
\end{table*}

\begin{figure}
    \centering
    \includegraphics[width=1\linewidth]{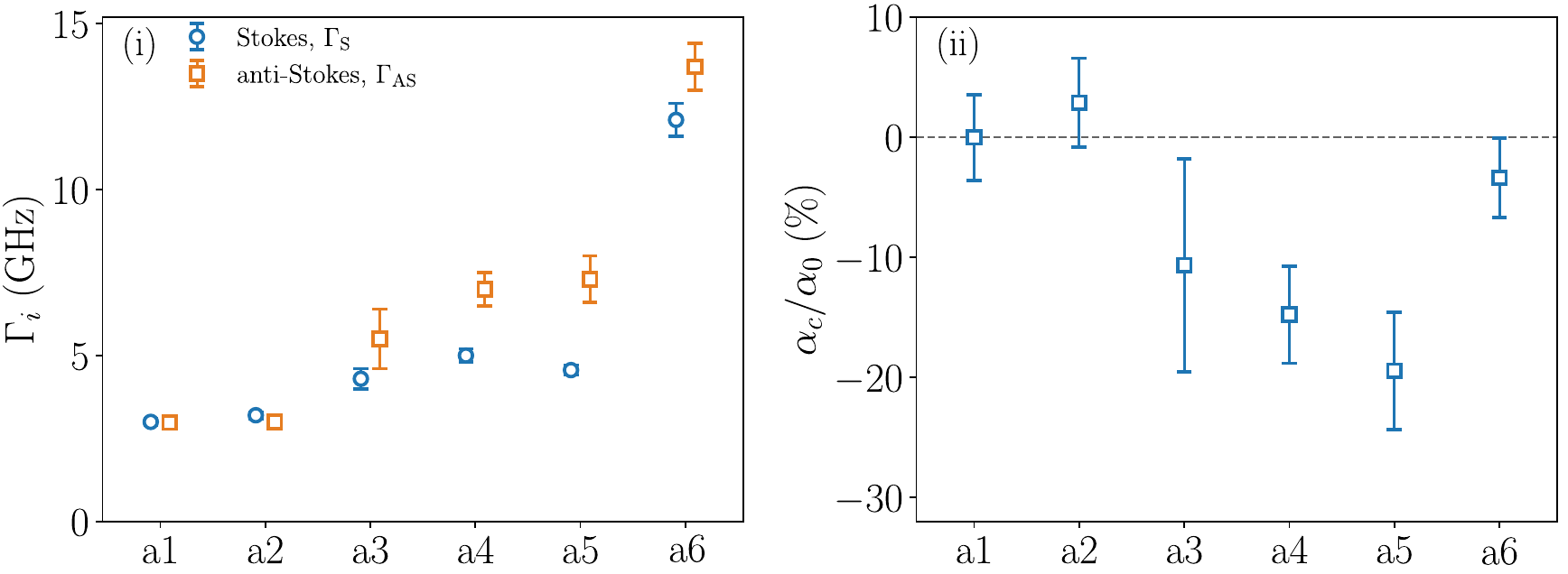}
    \caption{
(i) Stokes and anti-Stokes BLS linewidths, $\Gamma_{\mathrm{S}}$ and $\Gamma_{\mathrm{AS}}$, for samples a1-a5 and a6. (ii) Relative chiral-damping contribution $\alpha_c/\alpha_0$, obtained after removing the DMI-induced frequency contribution from the measured linewidth asymmetry according to Eq.~\eqref{eq:BLS_relative_chiral_damping}.
}
    \label{fig:placeholder}
\end{figure}

\section{Flow-regime characterization}
\label{sec:flow_characterization}

\noindent At sufficiently large driving fields, domain-wall motion is no longer governed by the thermally activated energy barriers characteristic of the creep regime \cite{metaxas_creep_2007}. The dynamics of chiral domain walls can instead be described within the two-dimensional framework developed for chiral domain walls \cite{ha_pham_very_2016}. Flow-regime measurements therefore provide a reference determination of the DMI that is substantially less sensitive to the pinning and elasticity effects that complicate the interpretation of asymmetric bubble expansion in the creep regime (see Main text).

Flow-regime measurements were performed on all Co-base samples described in Sec.~\ref{sec:samples_a1_a5} (i.e. samples a1-a5). For each sample, two complementary measurements were carried out. First, the DW velocity was measured as a function of the out-of-plane field $H_z$ in order to identify the characteristic Walker or high-velocity plateau regime. From the full velocity measurements (see Fig.~\ref{fig:walker_all_samples}), the depinning fields were extracted according to the procedure detailed in Ref.\cite{Jeudy2016}. Second, the domain-wall velocity was measured as a function of the in-plane field $H_x$ at a fixed out-of-plane driving field $H_z$ within the flow regime.  

\subsection{DMI constant from the Walker/plateau velocity}
\label{subsec:DMI constant from the Walker velocity}
\noindent A first and independent estimate of the DMI strength was obtained from the OOP field dependence of the domain-wall velocity plateaus. In systems with sufficiently strong DMI, the conventional velocity drop expected immediately above the Walker field is replaced by an extended high-velocity saturation field region. The characteristic Walker or plateau velocity is related to the DMI constant through \cite{ha_pham_very_2016,garcia_magnetic_2021}

\begin{equation}
    v_W=\frac{\pi}{2}\gamma\frac{D}{M_s},
    \label{eq:plateau}
\end{equation}

where $v_W$ is the characteristic Walker velocity, $\gamma$ is the gyromagnetic ratio, $D$ is the interfacial DMI energy density, and $M_s$ is the saturation magnetization.

\begin{figure*}[t]
\centering

\begin{subfigure}[t]{0.48\textwidth}
    \centering
    \includegraphics[width=\linewidth]
    {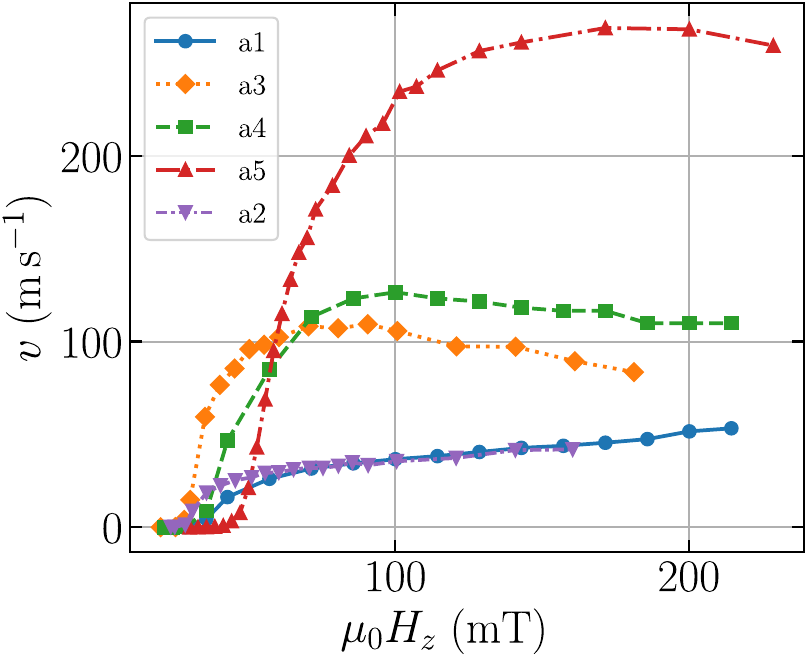}
    \label{fig:walker_velocity}
\end{subfigure}
\hfill
\begin{subfigure}[t]{0.48\textwidth}
    \centering
    \includegraphics[width=\linewidth]
    {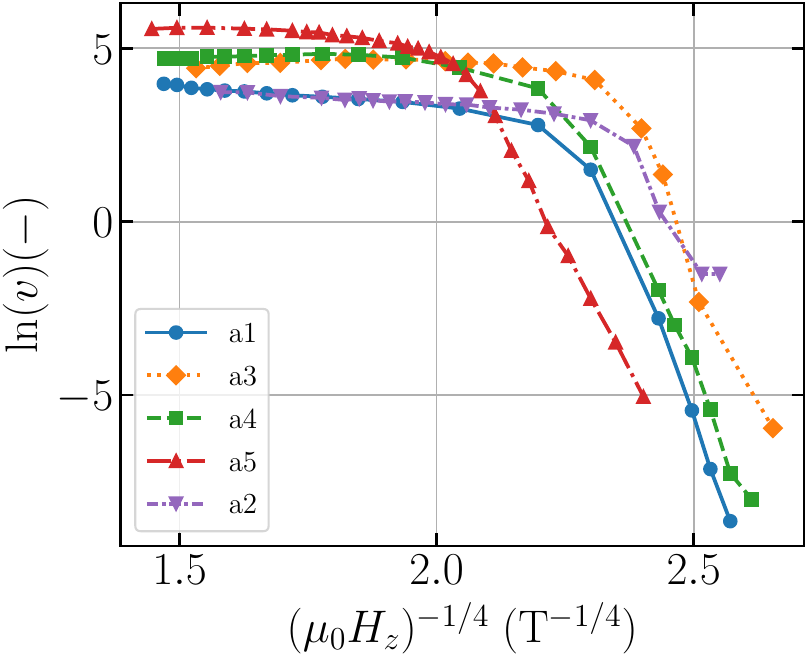}
    \label{fig:creep_scaling_all_samples}
\end{subfigure}

\vspace{0.5em}

\begin{subfigure}[t]{0.48\textwidth}
    \centering
    \includegraphics[width=\linewidth]
    {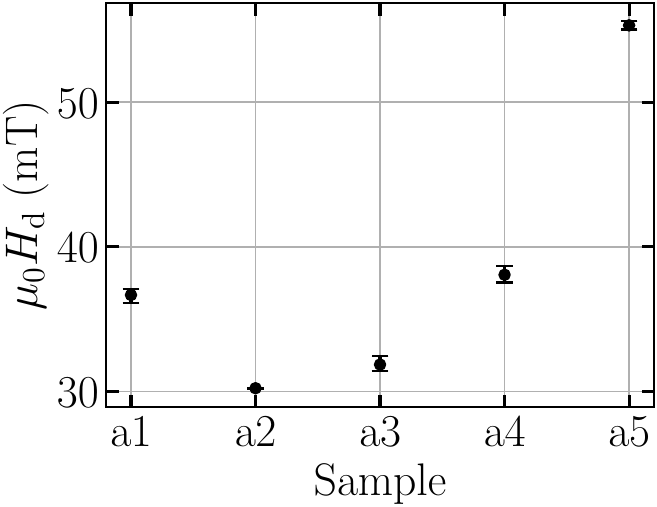}
    \label{fig:depinning_fields_all_samples}
\end{subfigure}

\caption{
Out-of-plane-field-dependent domain-wall dynamics the $a1-a5$ samples investigated in this work.
(a) Domain-wall velocity as a function of $\mu_0H_z$. The characteristic Walker or high-velocity plateau value $v_W$ is used to determine the DMI constant through Eq.~\eqref{eq:plateau}.
(b) Corresponding creep-law representation, $\ln v$ versus $(\mu_0H_z)^{-1/4}$.
(c) Depinning fields $\mu_0H_d$ extracted from the full creep-to-depinning \cite{Jeudy2016} fitting procedure for each sample.
}
\label{fig:walker_all_samples}

\end{figure*}

The domain-wall velocity was measured as a function of $\mu_0H_z$ for every sample in the a1-a5 series, allowing the corresponding characteristic velocity $v_W$ to be determined. The complete set of measurements is shown in Fig.~\ref{fig:walker_all_samples}. The DMI constants obtained using Eq.~\eqref{eq:plateau} are listed in Table~\ref{tab:Aex}.

The DMI constants obtained from the flow-regime velocity measurements are reported in Table~\ref{tab:Aex} and compared with the independently determined BLS values in Table~\ref{tab:DMI_comparison}. The general agreement between the two techniques supports the DMI values obtained in the creep-regime analysis presented in the main text.

\subsubsection{Creep characterization of the a6 sample}
\begin{figure}
    \centering
    \includegraphics[width=0.5\linewidth]{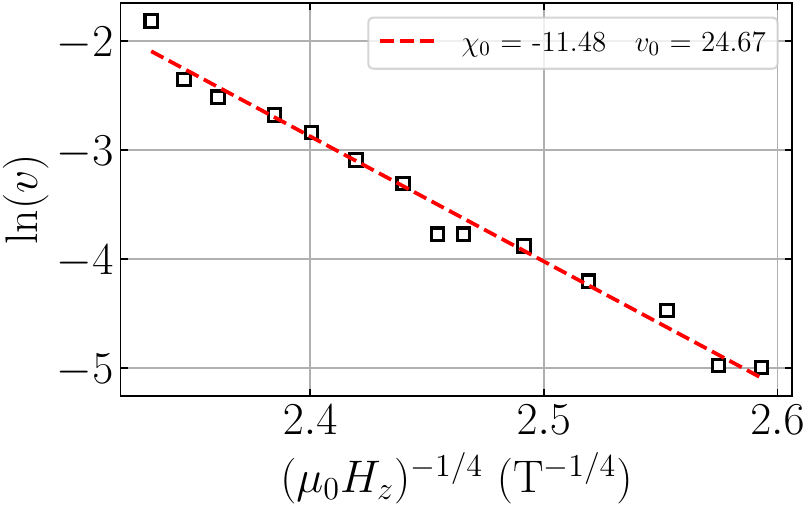}
    \caption{Out-of-plane-field-dependent domain-wall dynamics the a6 sample. The dashed line represents the linear fit of Eq.~\eqref{eq:creep_law}}.
    \label{fig:MAN60_creep}
\end{figure}

\noindent We were not able to measure the a6 sample up to the depinning regime as the same setup used for samples a1-a5 was unavailable. To check the validity of the creep regime, we performed low field bubble expansion measurements to confirm the linear scaling of $\ln(v)$ \cite{Lemerle1998}
\begin{equation}
    \ln(v) = v_0 + \chi_0 (\mu_0 H_z)^{-1/4} \label{eq:creep_law} 
\end{equation}
The results are shown in Fig.~\ref{fig:MAN60_creep}
\subsection{DMI field from in-plane-field compensation}
\label{subsec:DMI field from in-plane-field compensation}
\begin{figure*}[t]
    \centering

    \includegraphics[width=0.32\textwidth]
    {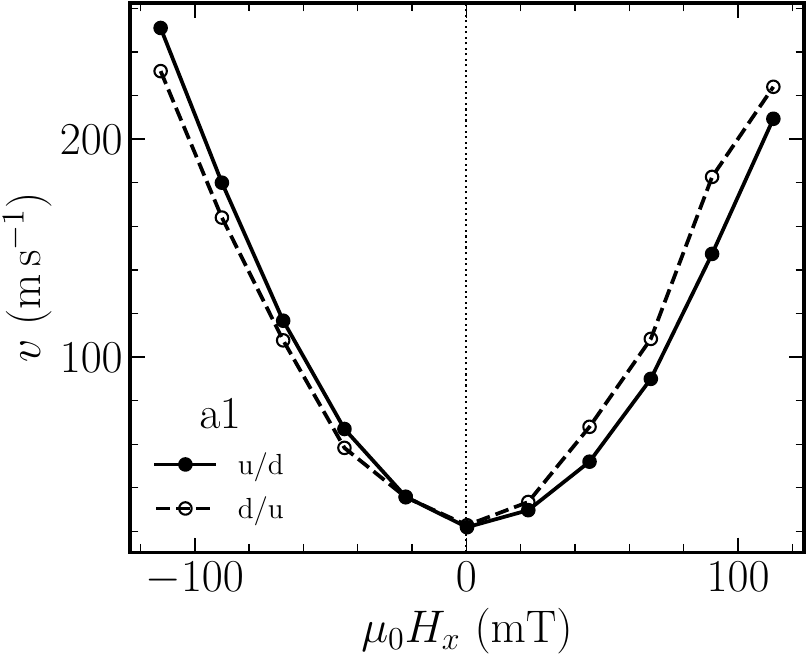}
    \hfill
    \includegraphics[width=0.32\textwidth]
    {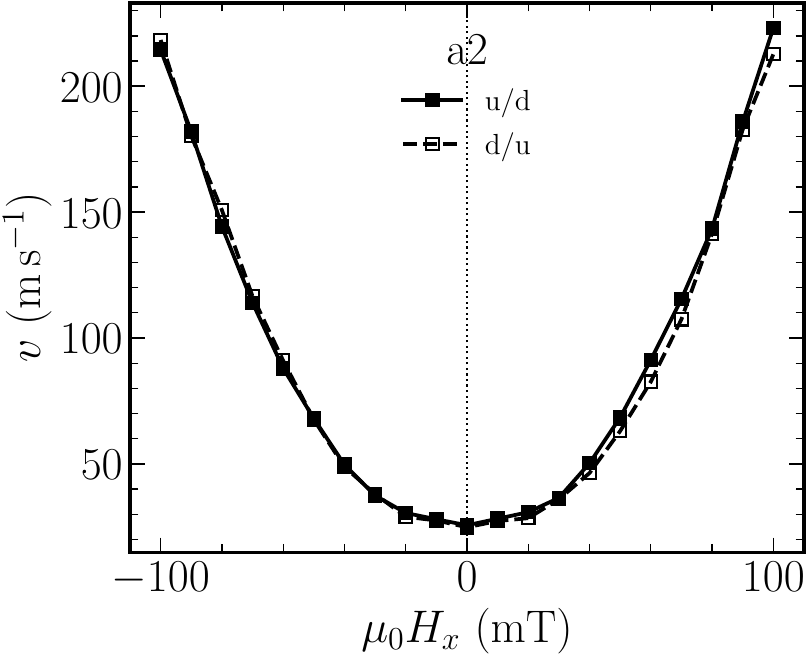}
    \hfill
    \includegraphics[width=0.32\textwidth]
    {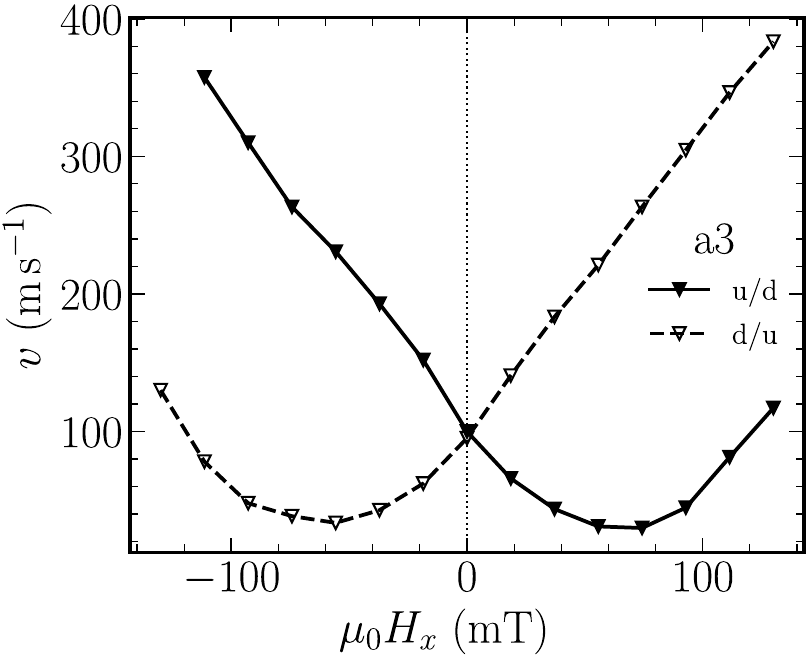}

    \vspace{0.5em}

    \includegraphics[width=0.32\textwidth]
    {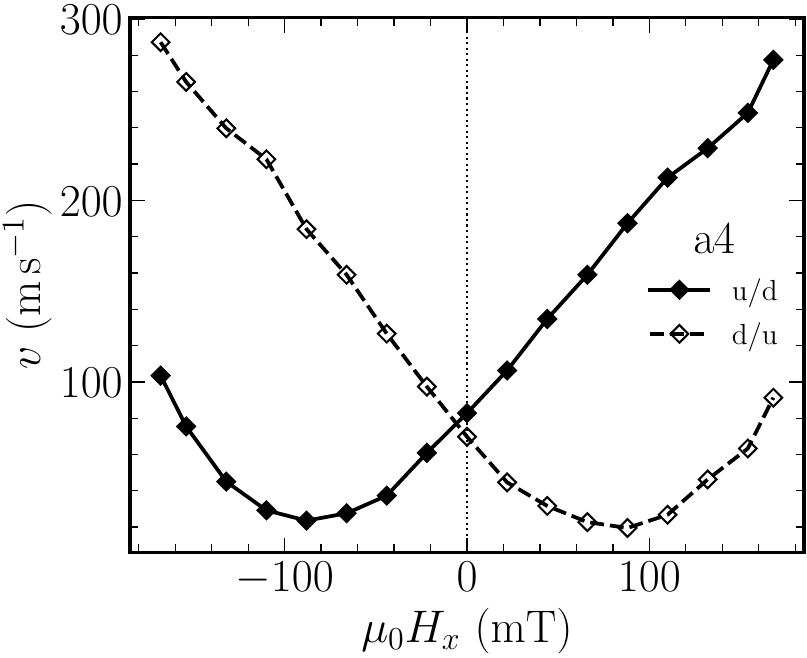}
    \hfill
    \includegraphics[width=0.32\textwidth]
    {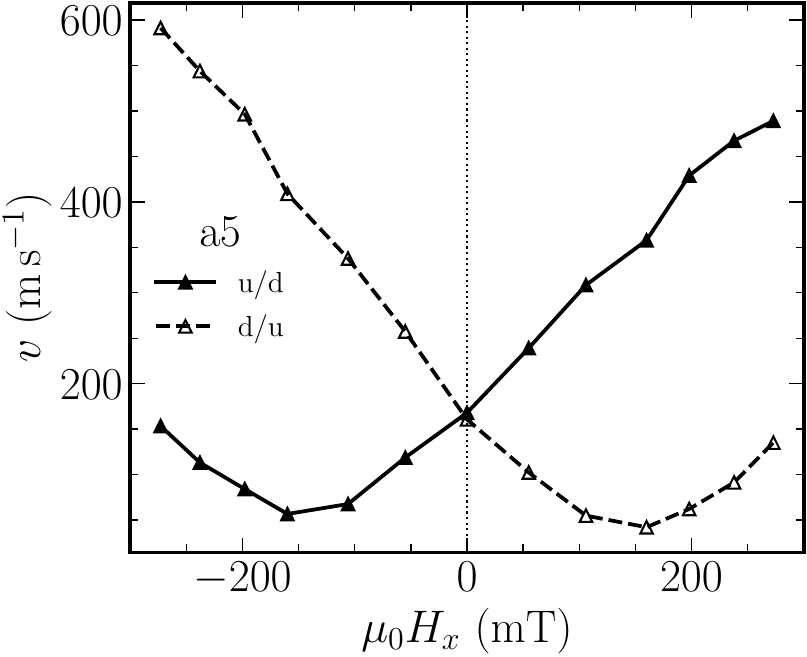}

    \caption{
    Domain-wall velocity measured as a function of the in-plane field
    $\mu_0H_x$ in the flow regime for samples $a_1$-$a_5$. For each
    sample, the velocities of the up-down (u/d) and down-up (d/u)
    portions of the expanding bubble domain are shown separately. The
    position of the velocity minimum identifies the compensation
    condition $\mu_0H_x=-\mu_0H_{\mathrm{DMI}} \equiv -\mu_0H_{\mathrm{DMI}}^{\mathrm{flow}}$
    \cite{krizakova_study_2019}.
    }
    \label{fig:flow_parabolas}
\end{figure*}

\noindent For each sample, asymmetric bubble expansion was measured as a function of the applied in-plane field $\mu_0H_x$, while maintaining the out-of-plane driving field within the flow regime (see Table.~\ref{tab:Aex}). The velocities of the up-down and down-up portions of the expanding bubble were evaluated independently. The velocity curves exhibit a minimum near the compensation condition at which the applied in-plane field balances the DMI effective field \cite{ha_pham_very_2016,krizakova_study_2019,garcia_magnetic_2021},
\begin{equation}
    H_x=-H_{\mathrm{DMI}}.
    \label{eq:flow_compensation}
\end{equation}
The flow-regime asymmetric bubble-expansion measurements for all samples are shown in Fig.~\ref{fig:flow_parabolas}. The recorded depinning fields $\mu_0 H_d$ for the a1-a5 samples and are reported in Table.~\ref{tab:Aex} alongside the recorded values of the effective DMI field $\mu_0H_{\mathrm{DMI}}$.

\subsection{Exchange-stiffness consistency check}

\noindent The two flow-regime measurements described above provide complementary information. The in-plane-field-dependent asymmetric bubble expansion determines the DMI effective field $H_{\mathrm{DMI}}$, whereas the out-of-plane-field-dependent velocity provides an independent estimate of the DMI energy density $D$. With the sign convention adopted in this work, the DMI constant and the corresponding effective field are related by

\begin{equation}
    D=\mu_0M_sH_{\mathrm{DMI}}\Delta,
    \label{eq:D_HD_relation}
\end{equation}
where

\begin{equation}
    \Delta=\sqrt{\frac{A}{K_{\mathrm{eff}}}}
    \label{eq:dw_width_flow}
\end{equation}
is the domain-wall-width parameter. Combining Eqs.~\eqref{eq:D_HD_relation} and \eqref{eq:dw_width_flow} gives

\begin{equation}
    A=
    K_{\mathrm{eff}}
    \left(
    \frac{D}{\mu_0M_sH_{\mathrm{DMI}}}
    \right)^2.
    \label{eq:A_from_flow}
\end{equation}

The extraction of $A$ is not the primary purpose of the flow measurements. Nevertheless, Eq.~\eqref{eq:A_from_flow} provides an independent consistency check on the material parameters entering the creep-regime calculations. This is particularly useful in ultrathin multilayers, where direct measurements of the exchange stiffness are experimentally challenging.
Conventional approaches determine the exchange stiffness $A$ from the temperature dependence of the saturation magnetization, using Bloch-law or related spin-wave analyses of $M_s(T)$ \cite{yastremsky_thermodynamics_2019,kuzmin_exchange_2020}, or from the exchange contribution to spin-wave spectra measured by ferromagnetic resonance, perpendicular standing spin-wave spectroscopy, or Brillouin light scattering \cite{devolder_exchange_2016,waring_exchange_2023}. In ultrathin multilayers, however, these methods can be affected by interface anisotropy, magnetic dead layers, proximity-induced magnetic moments, finite-size effects, and uncertain exchange boundary conditions. Applying Eq.~\eqref{eq:A_from_flow} to the samples of the a1-a5 series provides the exchange-stiffness estimates reported in Table~\ref{tab:Aex}.

\begin{table*}[htbp]
    \centering
    \caption{
    Applied OOP field $\mu_0 H_z$, DMI field $\mu_0 H_{DMI}^{\mathrm{flow}}$, depinning field $\mu_0 H_d$, DMI constant $D_{\mathrm{Walker}}$, and exchange stiffness obtained $A$ from
    flow-regime domain-wall velocity measurements. The DMI field is
    determined from the compensation minimum in the in-plane-field-dependent
    velocity. When a Walker or high-velocity plateau is observed, the DMI
    constant is obtained from Eq.~\eqref{eq:plateau}, and the exchange
    stiffness is subsequently calculated using Eq.~\eqref{eq:A_from_flow}.
    A relative uncertainty of $10\%$ is assigned to the graphically
    extracted values of $\mu_0 H_{DMI}^{\mathrm{flow}}$ and $D_{\mathrm{Walker}}$. The uncertainty
    in $A$ additionally includes the propagated uncertainties in $M_s$
    and $K_{\mathrm{eff}}$ (see Table.~\ref{tab:samples_Leeds}). No Walker plateau was observed for samples
    $a1$ and $a2$, preventing the extraction of $D_{\mathrm{Walker}}$ and $A$ by this
    procedure. No flow-regime measurements were performed on the a6
    sample. Here, n.o. denotes ``not observed'' and n.m. denotes
    ``not measured''.
    }
    \begin{tabular}{@{}llclcc@{}}
        \hline
        \hline
        Sample  &$\mu_0 H_z$&
        $\mu_0H_{\mathrm{DMI}}^{\mathrm{flow}}$ &$\mu_0 H_d$ &
        $D_\mathrm{Walker}
        $ &
        $A$ \\
         &(mT)&
        (mT)  &(mT)&
        (mJ/m$^2$) &
        (pJ/m) \\
        \hline
        $a1$   &$40$& $\approx 0$             &$36.7 \pm 0.5$& n.o.              & n.o. \\
        $a2$   &$40$& $\approx 0$             &$30.22 \pm 0.04$& n.o.              & n.o. \\
        $a3$   &$50$& $-70\pm 7$        &$31.9 \pm 0.5$& $-0.45\pm0.05$    & $13\pm4$ \\
        $a4$   &$50$& $-88\pm 9$        &$38.1 \pm 0.6 $& $-0.51\pm0.05$    & $12\pm4$ \\
        $a5$   &$100$& $-160\pm16$        &$55.3 \pm 0.3$& $-0.91\pm0.09$    & $23\pm6$ \\
        a6    &n.m.& n.m.                     &n.m.& n.m.               & n.m. \\
        \hline
        \hline
    \end{tabular}
    \label{tab:Aex}
\end{table*}

\subsection{Effect of chiral damping in the precessional regime}
\begin{figure}
 \centering
    \includegraphics[width=1\linewidth]{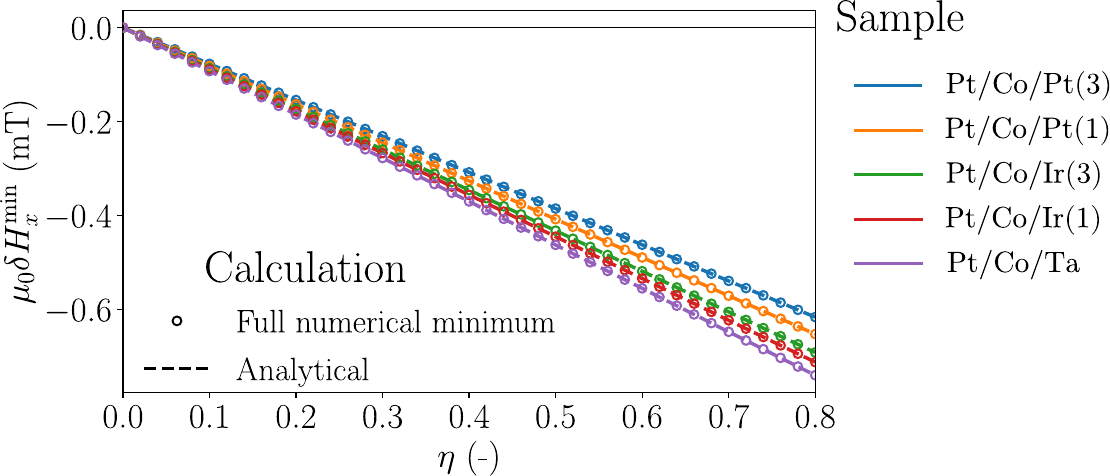}
    \caption{
    Difference between the minimum of the domain-wall velocity in the flow regime and the DMI effective field $\mu_0 \delta H_x^{\mathrm{min}} = \mu_0 [(H_x^\star)_{flow} + H_{\mathrm{DMI}}]$ as computed via the full numerical minimum of Eq.~\eqref{eq:prec_vel} and the analytical estimate Eq.~\eqref{eq:min_shift_flow}. The magnetic parameters used in the calculation are from the \(\mathrm{Pt/Co}/X\) series. We set Gilbert damping $\alpha_0 = 0.05$ and $\mu_0 H_z = 120$ mT for all samples. The selected $H_z$ field ensures correspondence with the precessional flow regime in all samples. See Tab.\ref{tab:Aex} for the depinning field analysis of the \(\mathrm{Pt/Co}/X\) series)
    }
    \label{fig:2_1}
\end{figure}

\noindent We consider the one-dimensional precessional dynamics with an angle-dependent damping
\begin{equation}
\alpha(\phi)=\alpha_0\left(1+\eta\cos\phi\right),
\qquad \bigg|\frac{\alpha_c}{\alpha_0}\bigg| \equiv  |\eta|<1.
\end{equation}
The deviation from the DMI compensation field is defined as
\begin{equation}
X=H_x+H_{\rm DMI},
\end{equation}
so that the compensation point corresponds to $X=0$. The internal torque entering the $q$-$\phi$ equations is written as
\begin{equation}
G(\phi,X)=\frac{H_K^{DW}}{2}\sin2\phi+\frac{\pi X}{2}\cos\phi.
\end{equation}
The cycle-averaged drift velocity is
\begin{equation}
\langle v\rangle=\gamma\Delta\frac{J}{I}, \label{eq:prec_vel}
\end{equation}
where
\begin{equation}
I=
\int_0^{2\pi}
\frac{1+\alpha(\phi)^2}
{H_z-\alpha(\phi)G(\phi,X)}
\,d\phi,
\end{equation}
and
\begin{equation}
J=
\int_0^{2\pi}
\frac{\alpha(\phi)H_z+G(\phi,X)}
{H_z-\alpha(\phi)G(\phi,X)}
\,d\phi.
\end{equation}
To perform the asymptotic expansion, we introduce the dimensionless parameters
\begin{equation}
\xi=\frac{\alpha_0H_K^{DW}}{2H_z},
\qquad
\rho=\frac{\pi X}{2H_z},
\end{equation}
so that
\begin{equation}
\frac{G(\phi,X)}{H_z}
=
\frac{\xi}{\alpha_0}\sin2\phi+\rho\cos\phi.
\end{equation}
Multiplying the denominator integral by $H_z$, we define $\bar I=H_zI$, giving
\begin{equation}
\langle v\rangle
=
\gamma\Delta H_z
\frac{J}{\bar I}.
\end{equation}
The integrands are expanded simultaneously in the small parameters $\rho$, $\eta$ and $\xi$. After integration over one full precessional cycle, all terms linear in $\xi$ vanish by symmetry, yielding
\begin{align}
\bar I =
2\pi(1+\alpha_0^2)
\Bigg[
&1+\frac{\xi^2}{2}
+
\frac{\alpha_0(1+3\alpha_0^2)}
{2(1+\alpha_0^2)}
\eta\rho
+
\frac{\alpha_0^2}{2}\rho^2 \nonumber
\\
&+
\frac{3\alpha_0(3+5\alpha_0^2)}
{4(1+\alpha_0^2)}
\eta\rho\xi^2
+
\frac{3\alpha_0^2}{2}\rho^2\xi^2
\
+
\mathcal O\left(
\eta^2,
\eta\rho^2,
\rho^3,
\xi^4
\right)
\Bigg],
\end{align}
and
\begin{align}
J
=
&2\pi\alpha_0
\Bigg[
1+
\frac{1+\alpha_0^2}
{2\alpha_0^2}\xi^2
+
\alpha_0\eta\rho
+
\frac{1+\alpha_0^2}{2}\rho^2 \nonumber
\\
&+
\frac{3(1+2\alpha_0^2)}
{2\alpha_0}
\eta\rho\xi^2
+
\frac{3(1+\alpha_0^2)}{2}\rho^2\xi^2
\
+
\mathcal{O} \left(
\eta^2,
\eta\rho^2,
\rho^3,
\xi^4
\right)
\Bigg].
\end{align}
Series division then gives
\begin{align}
\langle v\rangle
=
\gamma\Delta H_z
\frac{\alpha_0}{1+\alpha_0^2}
\Bigg[
1+
\frac{\xi^2}{2\alpha_0^2}
+
\frac{\rho^2}{2}
+
\rho^2\xi^2
\
+
\frac{\alpha_0(1-\alpha_0^2)}
{2(1+\alpha_0^2)}
\eta\rho
\\
+
\frac{5+5\alpha_0^2-2\alpha_0^4}
{4\alpha_0(1+\alpha_0^2)}
\eta\rho\xi^2
\
+
\mathcal O\left(
\eta^2,
\eta\rho^2,
\rho^3,
\xi^4
\right)
\Bigg].
\end{align}
The position of the extremum is obtained from
\begin{equation}
\frac{\partial\langle v\rangle}{\partial\rho}=0.
\end{equation}
Differentiating the expanded velocity gives
\begin{equation}
(1+2\xi^2)\rho_{\min}
+
\eta
\left[
\frac{\alpha_0(1-\alpha_0^2)}
{2(1+\alpha_0^2)}
+
\frac{5+5\alpha_0^2-2\alpha_0^4}
{4\alpha_0(1+\alpha_0^2)}
\xi^2
\right]
=0.
\end{equation}
which gives
\begin{equation}
\rho_{\min} =
-\eta
\frac{
\displaystyle
\frac{\alpha_0(1-\alpha_0^2)}
{2(1+\alpha_0^2)}
+
\frac{5+5\alpha_0^2-2\alpha_0^4}
{4\alpha_0(1+\alpha_0^2)}
\xi^2
}
{1+2\xi^2}.
\end{equation}
Since the numerator and denominator have been calculated only up to second order in $\xi$, their ratio must be consistently re-expanded to the same order. This yields
\begin{equation}
\rho_{\min}
-\eta
\left[
\frac{\alpha_0(1-\alpha_0^2)}
{2(1+\alpha_0^2)}
+
\frac{5+\alpha_0^2+2\alpha_0^4}
{4\alpha_0(1+\alpha_0^2)}
\xi^2
\right]
+
\mathcal O\left(
\eta^2,
\eta\xi^4
\right).
\end{equation}
Finally, using
\begin{equation}
\rho=\frac{\pi X}{2H_z},
\qquad
\xi=\frac{\alpha_0H_K^{DW}}{2H_z},
\end{equation}
the displacement of the minimum from the DMI compensation field is
\begin{equation}
\begin{aligned}
(H_x^\star)_{flow} 
&=
-H_{\rm DMI}
+
\frac{\alpha_0H_z}{\pi}
\frac{\eta}{1+\alpha_0^2}
\left[
1-\alpha_0^2
+
\frac{5+\alpha_0^2+2\alpha_0^4}{8}
\left(
\frac{H_K^{DW}}{H_z}
\right)^2
\right]
+
\mathcal O.
\end{aligned}
\end{equation}
For $\alpha_0\ll1$, this simplifies to the expressions used in the main text (Eq.~(17))
\begin{equation}
(H_x^\star)_{flow} 
\simeq -H_{\rm DMI}
+
\frac{\eta\alpha_0H_z}{\pi}
\left[
1+
\frac{5}{8}
\left(
\frac{H_K^{DW}}{H_z}
\right)^2
\right].
\label{eq:min_shift_flow}
\end{equation}
In Fig.\ref{fig:2_1} we report both the minimum obtained numerically from Eq.\eqref{eq:prec_vel} and the analytical estimate of Eq.\eqref{eq:min_shift_flow}.

\section{Derivation of the three-amplitude model}
\label{sec:SI_three_amplitude_derivation}

\noindent In this section, we derive the symmetry-decomposed model used in the main
text. We start from a creep-velocity Ansatz of the form
\begin{equation}
v(H_x,H_z)
=
v_\star
\frac{\sigma_{\rm eq}(H_x)^\beta}
{\alpha_{\rm eff}(H_x)}
\exp\left[
-\chi_0
\left(
\frac{\tilde{\sigma}(H_x)}
{\tilde{\sigma}(0)}
\right)^{1/4}
H_z^{-1/4}
\right].
\label{eq:SI_creep_ansatz}
\end{equation}
Here, \(v_\star\) is a field-independent attempt-frequency prefactor,
\(\sigma_{\rm eq}(H_x)\) is the equilibrium domain-wall energy density,
and \(\tilde{\sigma}(H_x)\) is the domain-wall stiffness in the
long-wavelength limit \cite{pellegren_dispersive_2017,lau_disentangling_2018}.
\begin{equation}
\tilde{\sigma}(\theta)
=
\sigma_{\rm eq}(\theta)
+
\frac{\partial^2\sigma_{\rm eq}}{\partial\theta^2}(\theta),
\qquad
\sigma_{\rm eq}(\theta)
=
\min_\phi \sigma(\theta,\phi).
\label{eq:SI_stiffness}
\end{equation}
The quantity \(\chi_0\) is the creep scaling parameter at \(H_x=0\),
whereas \(\beta\) is an effective exponent describing the dependence of
the velocity prefactor on the equilibrium wall energy
\cite{lau_disentangling_2018}. It should not be confused with the reduced
in-plane field \(\beta_x\) introduced below.
Taking the natural logarithm of the velocity normalized at \(H_x=0\)
gives
\begin{align}
\ln\left[
\frac{v(H_x,H_z)}{v(0,H_z)}
\right]
&=
\beta
\ln\left[
\frac{\sigma_{\rm eq}(H_x)}
{\sigma_{\rm eq}(0)}
\right]
+
\ln\left[
\frac{\alpha_{\rm eff}(0)}
{\alpha_{\rm eff}(H_x)}
\right]
\nonumber\\
&\quad
-
\chi_0H_z^{-1/4}
\left[
\left(
\frac{\tilde{\sigma}(H_x)}
{\tilde{\sigma}(0)}
\right)^{1/4}
-1
\right].
\label{eq:SI_lnv_exact}
\end{align}
For compactness, we define the relative variations
\begin{equation}
\delta\sigma_{\rm eq}(H_x)
\equiv
\frac{\sigma_{\rm eq}(H_x)-\sigma_{\rm eq}(0)}
{\sigma_{\rm eq}(0)},
\qquad
\delta\tilde{\sigma}(H_x)
\equiv
\frac{\tilde{\sigma}(H_x)-\tilde{\sigma}(0)}
{\tilde{\sigma}(0)}.
\label{eq:SI_relative_variations}
\end{equation}
Equation~\eqref{eq:SI_lnv_exact} can then be expressed as
\begin{align}
\ln\left[
\frac{v(H_x,H_z)}{v(0,H_z)}
\right]
&=
\beta
\ln\left[
1+\delta\sigma_{\rm eq}(H_x)
\right]
+
\ln\left[
\frac{\alpha_{\rm eff}(0)}
{\alpha_{\rm eff}(H_x)}
\right]
\nonumber\\
&\quad
-
\chi_0
\left[
\left(
1+\delta\tilde{\sigma}(H_x)
\right)^{1/4}
-1
\right]
H_z^{-1/4}.
\label{eq:SI_lnv_variations}
\end{align}
 Considering the experimentally accessible IP field ranges is asymmetric bubble domain expansions ($|\mu_0 H_x| \le 200$ mT) we assume small in-plane-field-induced variations,
\begin{equation}
\left|\delta\sigma_{\rm eq}(H_x)\right|\ll1,
\qquad
\left|\delta\tilde{\sigma}(H_x)\right|\ll1,
\end{equation}
Since direct evaluation of \(\widetilde{\sigma}(H_x)/\widetilde{\sigma}(0)\), accounting for both mixed and saturated zero-field configurations, shows that the first-order fourth-root expansion is accurate to better than \(0.3\%\) for samples a2–a5 over their respective experimental field ranges. Sample a1 provides the most restrictive case, with a maximum error of approximately 3.6 \%. With the mentioned assumption, we obtain
\begin{align}
\ln\left[
\frac{v(H_x,H_z)}{v(0,H_z)}
\right]
&\simeq
\beta
\left[
\frac{\sigma_{\rm eq}(H_x)}
{\sigma_{\rm eq}(0)}
-1
\right]
+
\ln\left[
\frac{\alpha_{\rm eff}(0)}
{\alpha_{\rm eff}(H_x)}
\right]
\nonumber\\
&\quad
-
C_z
\left[
\frac{\tilde{\sigma}(H_x)}
{\tilde{\sigma}(0)}
-1
\right],
\label{eq:SI_lnv_linearized}
\end{align}
where
\begin{equation}
C_z
=
\frac{\chi_0H_z^{-1/4}}{4}.
\label{eq:SI_Cz}
\end{equation}
To retain the correct normalizations of the two
energy-dependent terms, we define
\begin{equation}
S_0
\equiv
\frac{\sigma_{\rm eq}(0)}{\sigma_0},
\qquad
T_0
\equiv
\frac{\tilde{\sigma}(0)}{\sigma_0},
\label{eq:SI_S0T0}
\end{equation}
where
\begin{equation}
\sigma_0=4\sqrt{AK_{\rm eff}}
\end{equation}
is the Bloch-wall energy density. Equation~\eqref{eq:SI_lnv_linearized}
then becomes
\begin{align}
\ln\left[
\frac{v(H_x,H_z)}{v(0,H_z)}
\right]
&\simeq
\beta
\left[
\frac{\sigma_{\rm eq}(H_x)/\sigma_0}{S_0}
-1
\right]
+
\ln\left[
\frac{\alpha_{\rm eff}(0)}
{\alpha_{\rm eff}(H_x)}
\right]
\nonumber\\
&\quad
-
C_z
\left[
\frac{\tilde{\sigma}(H_x)/\sigma_0}{T_0}
-1
\right].
\label{eq:SI_lnv_ST}
\end{align}
The two reference values \(S_0\) and \(T_0\) are generally different:
the prefactor contribution is normalized by the equilibrium wall energy
at \(H_x=0\), whereas the activation-barrier contribution is normalized
by the stiffness at \(H_x=0\). We introduce the dimensionless quantities
\begin{equation}
\zeta
=
\frac{\pi\mu_0M_s\Delta H_{\rm DMI}}{\sigma_0},
\qquad
\beta_x
=
\frac{\pi\mu_0M_s\Delta H_x}{\sigma_0},
\qquad
\kappa
=
\frac{\pi\mu_0M_s\Delta H_K^{\rm DW}}{2\sigma_0}.
\label{eq:SI_reduced_parameters}
\end{equation}
In terms of these quantities, the domain-wall energy density is
\begin{equation}
\sigma(\theta,\phi)
=
\sigma_0
\left[
1
-
\zeta\cos(\phi-\theta)
-
\beta_x\cos\phi
+
\kappa\cos^2(\phi-\theta)
\right].
\label{eq:SI_dw_energy_reduced}
\end{equation}
For notational compactness, we introduce
\begin{equation}
X
=
H_x+H_{\rm DMI}.
\label{eq:SI_shifted_field}
\end{equation}
The reference axis for all the subsequent symmetry decompositions of the velocity curve is therefore \(X=0\), while the
mixed Bloch-N\'eel region corresponds to
\[
|X|\leq H_K^{\rm DW}.
\]
For the right-hand side of the expanding bubble, we set \(\theta=0\).
Minimizing Eq.~\eqref{eq:SI_dw_energy_reduced} with respect to \(\phi\)
gives
\begin{equation}
\cos\phi_{\rm eq}
=
\begin{cases}
-1,
& X<-H_K^{\rm DW},
\\[4pt]
\dfrac{X}{H_K^{\rm DW}},
& |X|\leq H_K^{\rm DW},
\\[8pt]
1,
& X>H_K^{\rm DW}.
\end{cases}
\label{eq:SI_phi_eq}
\end{equation}
The corresponding equilibrium wall energy is
\begin{equation}
\frac{\sigma_{\rm eq}(X)}{\sigma_0}
=
\begin{cases}
1+\kappa
-
2\kappa\dfrac{|X|}{H_K^{\rm DW}},
& |X|>H_K^{\rm DW},
\\[10pt]
1-\kappa
\left(
\dfrac{X}{H_K^{\rm DW}}
\right)^2,
& |X|\leq H_K^{\rm DW}.
\end{cases}
\label{eq:SI_sigma_eq}
\end{equation}
It is therefore symmetric around the compensation axis:
\begin{equation}
\sigma_{\rm eq}(X)
=
\sigma_{\rm eq}(-X).
\label{eq:SI_sigma_symmetry}
\end{equation}
Consequently, the equilibrium-energy contribution to the velocity
prefactor contributes only to the symmetric part of the creep curve. Using Eq.~\eqref{eq:SI_stiffness}, the symmetric part of the stiffness is
\begin{equation}
\frac{\tilde{\sigma}_S(X)}{\sigma_0}
=
\begin{cases}
1-\kappa
-
2\kappa\dfrac{|X|}{H_K^{\rm DW}},
& |X|>H_K^{\rm DW},
\\[10pt]
1+2\kappa
-
5\kappa
\left(
\dfrac{X}{H_K^{\rm DW}}
\right)^2,
& |X|\leq H_K^{\rm DW},
\end{cases}
\label{eq:SI_stiffness_symmetric}
\end{equation}
whereas its antisymmetric part is
\begin{equation}
\frac{\tilde{\sigma}_A(X)}{\sigma_0}
=
\begin{cases}
\zeta\,{\rm sgn}(X),
& |X|>H_K^{\rm DW},
\\[8pt]
\zeta
\dfrac{X}{H_K^{\rm DW}},
& |X|\leq H_K^{\rm DW}.
\end{cases}
\label{eq:SI_stiffness_antisymmetric}
\end{equation}
The normalization point \(H_x=0\) corresponds to \(X=H_{\rm DMI}\).
The normalization constants are therefore obtained by evaluating the
full equilibrium energy and stiffness at this point:
\begin{equation}
S_0
=
\left.
\frac{\sigma_{\rm eq}(X)}{\sigma_0}
\right|_{X=H_{\rm DMI}},
\qquad
T_0
=
\left.
\frac{\tilde{\sigma}(X)}{\sigma_0}
\right|_{X=H_{\rm DMI}}.
\label{eq:SI_S0T0_evaluation}
\end{equation}
Explicitly,
\begin{equation}
S_0
=
\begin{cases}
1+\kappa
-
2\kappa\dfrac{|H_{\rm DMI}|}{H_K^{\rm DW}},
& |H_{\rm DMI}|>H_K^{\rm DW},
\\[10pt]
1-\kappa
\left(
\dfrac{H_{\rm DMI}}{H_K^{\rm DW}}
\right)^2,
& |H_{\rm DMI}|\leq H_K^{\rm DW},
\end{cases}
\label{eq:SI_S0_explicit}
\end{equation}
and
\begin{equation}
T_0
=
\begin{cases}
1-\kappa
-
2\kappa\dfrac{|H_{\rm DMI}|}{H_K^{\rm DW}}
+
\zeta\,{\rm sgn}(H_{\rm DMI}),
& |H_{\rm DMI}|>H_K^{\rm DW},
\\[10pt]
1+2\kappa
-
5\kappa
\left(
\dfrac{H_{\rm DMI}}{H_K^{\rm DW}}
\right)^2
+
\zeta
\dfrac{H_{\rm DMI}}{H_K^{\rm DW}},
& |H_{\rm DMI}|\leq H_K^{\rm DW}.
\end{cases}
\label{eq:SI_T0_explicit}
\end{equation}
The effective damping is written as
\begin{equation}
\alpha_{\rm eff}(\phi)
=
\alpha_0
\left[
1+s\eta\cos\phi
\right],
\label{eq:SI_alpha_eff}
\end{equation}
where \(s=\pm1\) labels the two domain-wall chiralities and \(\eta\) is
the relative chiral-damping strength
\cite{akosa_phenomenology_2016,akosa_decoupling_2024}. Expanding the
logarithmic damping contribution to second order in \(\eta\) gives
\begin{align}
\ln\left[
\frac{\alpha_{\rm eff}(0)}
{\alpha_{\rm eff}(X)}
\right]
&=
{\rm const.}
-
s\eta\cos\phi_{\rm eq}
+
\frac{\eta^2}{2}
\cos^2\phi_{\rm eq}
+
O(\eta^3).
\label{eq:SI_damping_expansion}
\end{align}
The first-order term is antisymmetric in \(X\), whereas the second-order
term is symmetric. Since \(s=\pm1\) is a convention-dependent chirality
label, we absorb the minus sign in the linear term into the definition of
\(s\) from this point onward. With the convention used in the main text,
the linear contribution to the antisymmetric amplitude is therefore
written as \(s\eta\). Collecting all symmetric contributions, and removing one common
field-independent offset, symmetric part of the function can be
written as
\begin{equation}
(\ln v)_S(X)
=
\begin{cases}
\nu_1(H_K^{\rm DW})^2
-
\nu_3\left(X+H_K^{\rm DW}\right),
& X<-H_K^{\rm DW},
\\[6pt]
\nu_1X^2,
& |X|\leq H_K^{\rm DW},
\\[6pt]
\nu_1(H_K^{\rm DW})^2
+
\nu_3\left(X-H_K^{\rm DW}\right),
& X>H_K^{\rm DW}.
\end{cases}
\label{eq:SI_lnv_symmetric}
\end{equation}
The antisymmetric part is
\begin{equation}
(\ln v)_A(X)
=
\begin{cases}
-\nu_2,
& X<-H_K^{\rm DW},
\\[6pt]
\nu_2\dfrac{X}{H_K^{\rm DW}},
& |X|\leq H_K^{\rm DW},
\\[10pt]
\nu_2,
& X>H_K^{\rm DW}.
\end{cases}
\label{eq:SI_lnv_antisymmetric}
\end{equation}
The three amplitudes appearing in
Eqs.~\eqref{eq:SI_lnv_symmetric} and
\eqref{eq:SI_lnv_antisymmetric} are
\begin{align}
\nu_1
&=
\frac{1}{(H_K^{\rm DW})^2}
\left[
-\frac{\beta\kappa}{S_0}
+
\frac{5C_z\kappa}{T_0}
+
\frac{\eta^2}{2}
\right],
\label{eq:SI_nu1}
\\
\nu_2
&=
s\eta
-
\frac{\zeta C_z}{T_0},
\label{eq:SI_nu2}
\\
\nu_3
&=
\frac{2\kappa}{H_K^{\rm DW}}
\left[
\frac{C_z}{T_0}
-
\frac{\beta}{S_0}
\right].
\label{eq:SI_nu3}
\end{align}
Thus, \(\nu_1\) has units of T\(^{-2}\), \(\nu_2\) is
dimensionless, and \(\nu_3\) has units of T\(^{-1}\). Dimensionless versions of the amplitudes can be derived by multiplying the values obtained from the fit by the DW anisotropy field $\bar\nu_1 = (H_K^{DW})^2 \nu_1 \, ,\bar\nu_2 = \nu_2 \, , \,\bar\nu_3 =  H_K^{DW} \nu_3 $. A complete  set of the dimensionless fitting amplitudes $\bar\nu_{1 \cdots 3}$ is reported in Table.~\ref{tab:three_amplitude_parameters}.
Equations~\eqref{eq:SI_lnv_symmetric} and
\eqref{eq:SI_lnv_antisymmetric} define the symmetric and antisymmetric
shape functions about the compensation axis $X=0$, with their vertical
origin chosen such that both shape functions vanish at $X=0$. No
independent vertical-offset parameter is introduced in the fit.

The experimental and calculated velocity curves are instead normalized
at $H_x=0$. Defining the total symmetry-centered shape as

\begin{equation}
F(X)
=
(\ln v)_S(X)
+
(\ln v)_A(X),
\label{eq:SI_total_shape}
\end{equation}
the function compared with the normalized experimental velocity is
\begin{equation}
\ln\left[
\frac{v(H_x,H_z)}
{v(0,H_z)}
\right]_{\rm model}
=
F\left(H_x+H_{\rm DMI}\right)
-
F\left(H_{\rm DMI}\right).
\label{eq:SI_model_normalization}
\end{equation}
Indeed, at the experimental normalization point $H_x=0$, the shifted
field is $X=H_{\rm DMI}$, and therefore
\begin{equation}
\left.
\ln\left[
\frac{v(H_x,H_z)}
{v(0,H_z)}
\right]_{\rm model}
\right|_{H_x=0}
=
F(H_{\rm DMI})-F(H_{\rm DMI})
=
0.
\label{eq:SI_model_zero}
\end{equation}
The term $F(H_{\rm DMI})$ is a single parameter-dependent constant
evaluated at the normalization point. It is recalculated for each trial
set of fitting parameters, but it does not constitute an additional fit
parameter. In particular, it does not modify the amplitudes $\nu_1$,
$\nu_2$, or $\nu_3$: it only fixes the vertical origin of the complete
model.
Since a field-independent constant is symmetric under the transformation
$X\rightarrow -X$, the strict symmetry decomposition of the normalized
curve may equivalently be written as
\begin{align}
\left(\ln\frac{v}{v(0)}\right)_S
&=
(\ln v)_S(X)
-
F(H_{\rm DMI}),
\label{eq:SI_normalized_symmetric}
\\
\left(\ln\frac{v}{v(0)}\right)_A
&=
(\ln v)_A(X).
\label{eq:SI_normalized_antisymmetric}
\end{align}
Thus, the normalization affects only the constant level of the symmetric
component. It does not alter the symmetric curvature $\nu_1$, the high-field
slope $\nu_3$, the antisymmetric amplitude $\nu_2$, or the position of the velocity
minimum. In the mixed Bloch-N\'eel region, the field-dependent part of the
velocity is
\begin{equation}
\ln\left[
\frac{v(H_x,H_z)}{v(0,H_z)}
\right]
=
{\rm const.}
+
\nu_1X^2
+
\nu_2\frac{X}{H_K^{\rm DW}},
\qquad
X=H_x+H_{\rm DMI}.
\label{eq:SI_mixed_branch}
\end{equation}
The apparent minimum is therefore located at
\begin{equation}
(H_x^\star)_{creep}
=
-H_{\rm DMI}
-
\frac{\nu_2}
{2\nu_1H_K^{\rm DW}}.
\label{eq:SI_apparent_minimum}
\end{equation}
Equation~\eqref{eq:SI_apparent_minimum} applies provided that the
stationary point remains inside the mixed Bloch-N\'eel region and is a
minimum:
\begin{equation}
\left|
\frac{\nu_2}{2\nu_1H_K^{\rm DW}}
\right|
\leq H_K^{\rm DW},
\qquad
\nu_1>0.
\label{eq:SI_minimum_conditions}
\end{equation}

\subsection{Normalization regimes}
\label{sec:SI_normalization_regimes}
When the normalization point \(H_x=0\) lies in either saturated
N\'eel branch, \(|\zeta|>2\kappa\), one has
\begin{equation}
S_0
=
1+\kappa-|\zeta|,
\qquad
T_0
=
1-\kappa.
\label{eq:SI_normalization_saturated}
\end{equation}
For the positive saturated branch, \(\zeta>2\kappa\), this reduces to
\(S_0=1-\zeta+\kappa\). The corresponding coefficients are
\begin{align}
\nu_1(H_K^{\rm DW})^2
&=
-\frac{\beta\kappa}
{1+\kappa-|\zeta|}
+
\frac{5C_z\kappa}
{1-\kappa}
+
\frac{\eta^2}{2},
\label{eq:SI_nu1_saturated}
\\
\nu_2
&=
s\eta
-
\frac{\zeta C_z}
{1-\kappa},
\label{eq:SI_nu2_saturated}
\\
\nu_3H_K^{\rm DW}
&=
2\kappa
\left[
\frac{C_z}{1-\kappa}
-
\frac{\beta}{1+\kappa-|\zeta|}
\right].
\label{eq:SI_nu3_saturated}
\end{align}
If instead the normalization point lies inside the mixed Bloch-N\'eel
region, \(|\zeta|\leq2\kappa\), then
\begin{equation}
S_0
=
1-\frac{\zeta^2}{4\kappa},
\qquad
T_0
=
1+2\kappa-\frac{3\zeta^2}{4\kappa}.
\label{eq:SI_normalization_mixed}
\end{equation}
The coefficients become
\begin{align}
\nu_1(H_K^{\rm DW})^2
&=
-\frac{\beta\kappa}
{1-\dfrac{\zeta^2}{4\kappa}}
+
\frac{5C_z\kappa}
{1+2\kappa-\dfrac{3\zeta^2}{4\kappa}}
+
\frac{\eta^2}{2},
\label{eq:SI_nu1_mixed}
\\
\nu_2
&=
s\eta
-
\frac{\zeta C_z}
{1+2\kappa-\dfrac{3\zeta^2}{4\kappa}},
\label{eq:SI_nu2_mixed}
\\
\nu_3H_K^{\rm DW}
&=
2\kappa
\left[
\frac{C_z}
{1+2\kappa-\dfrac{3\zeta^2}{4\kappa}}
-
\frac{\beta}
{1-\dfrac{\zeta^2}{4\kappa}}
\right].
\label{eq:SI_nu3_mixed}
\end{align}

\subsection{Additional fits and full dataset}
\begin{figure}
    \centering
    \includegraphics[width=0.9\linewidth]{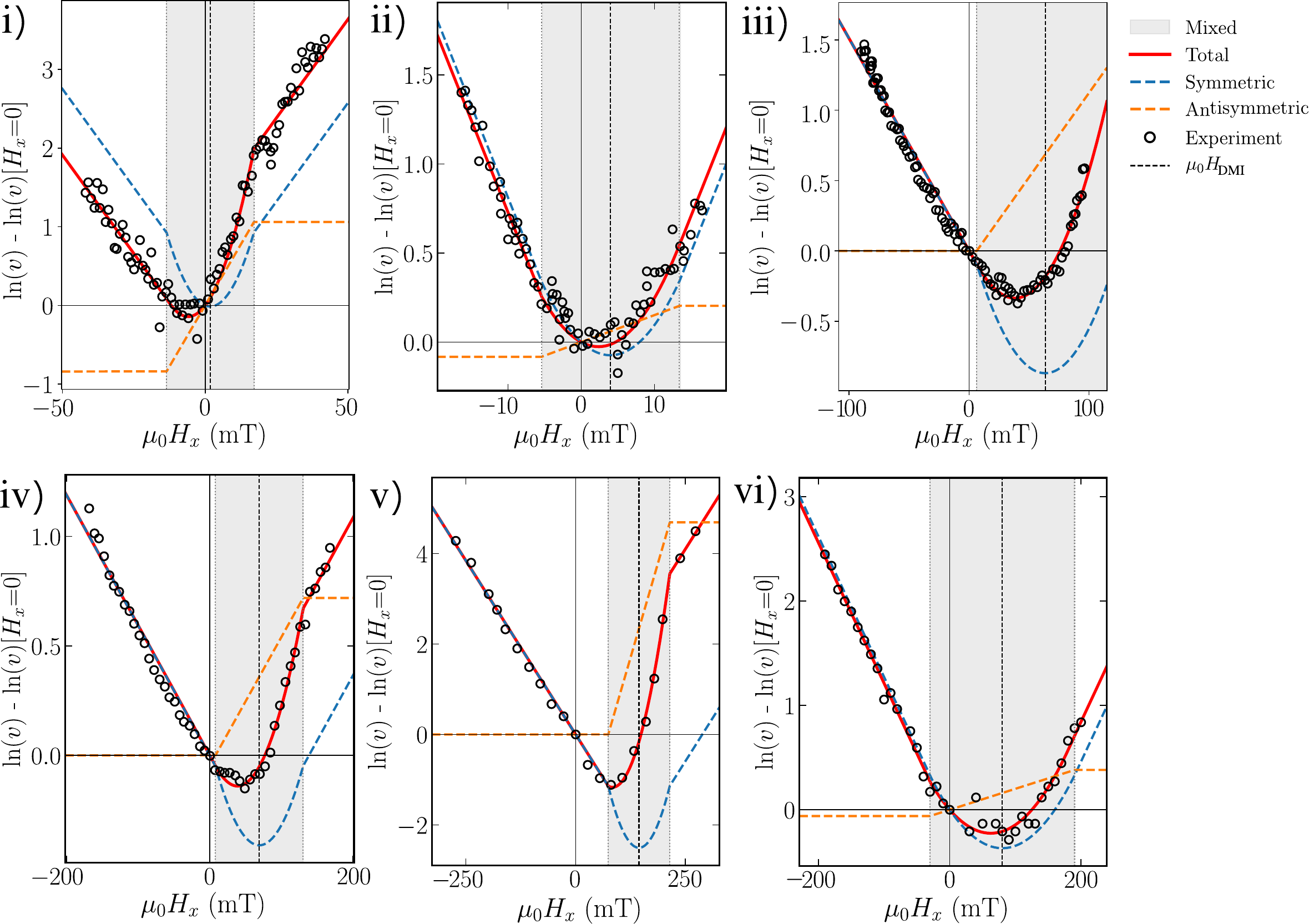}
    \caption{Complete sample series fits with the three amplitude model of Eqs.~(6-7) performed on the normalized velocity $\ln(v(H_x)/v(0))$. Panels i)-v) show the fits for samples a1-a5 while panel vi) shows the fit of sample a6. The red line represent the full fitted curve, while the blue and orange dashed lines represent the symmetric and antisymmetric components respectively. The vertical dashed line represents the fitted $\mu_0 H_{DMI}^{\mathrm{new}}$ field reported in Table.~\ref{tab:three_amplitude_parameters}. The shaded region displays the mixed Bloch-Néel transition region.}
    \label{fig:full_fits}
\end{figure}

\noindent
The complete three-amplitude fits for all investigated samples are shown in
Fig.~\ref{fig:full_fits}, while the corresponding fitting parameters are
summarized in Table~\ref{tab:three_amplitude_parameters}. The parameter $H_K^{DW}$
is treated as an effective crossover field delimiting the central mixed
Bloch-Néel region, rather than as a direct measurement of the microscopic
domain-wall anisotropy field. Its extraction is most reliable when the
experimental range samples both crossover boundaries,
$H_x=-H_{\mathrm{DMI}}\pm H_K^{DW}$, together with sufficient data in the outer
branches. When one or both boundaries lie outside the accessible field range,
$H_K^{DW}$ becomes correlated with $H_{\mathrm{DMI}}$ and the fitted amplitudes and
is consequently more sensitive to the imposed bounds. In these cases,
independently calculated magnetic parameters provide a useful physical
constraint, but the resulting DMI field should be interpreted in light of the limited transition coverage. 

\begin{figure}
    \centering
    \includegraphics[width=0.7\linewidth]{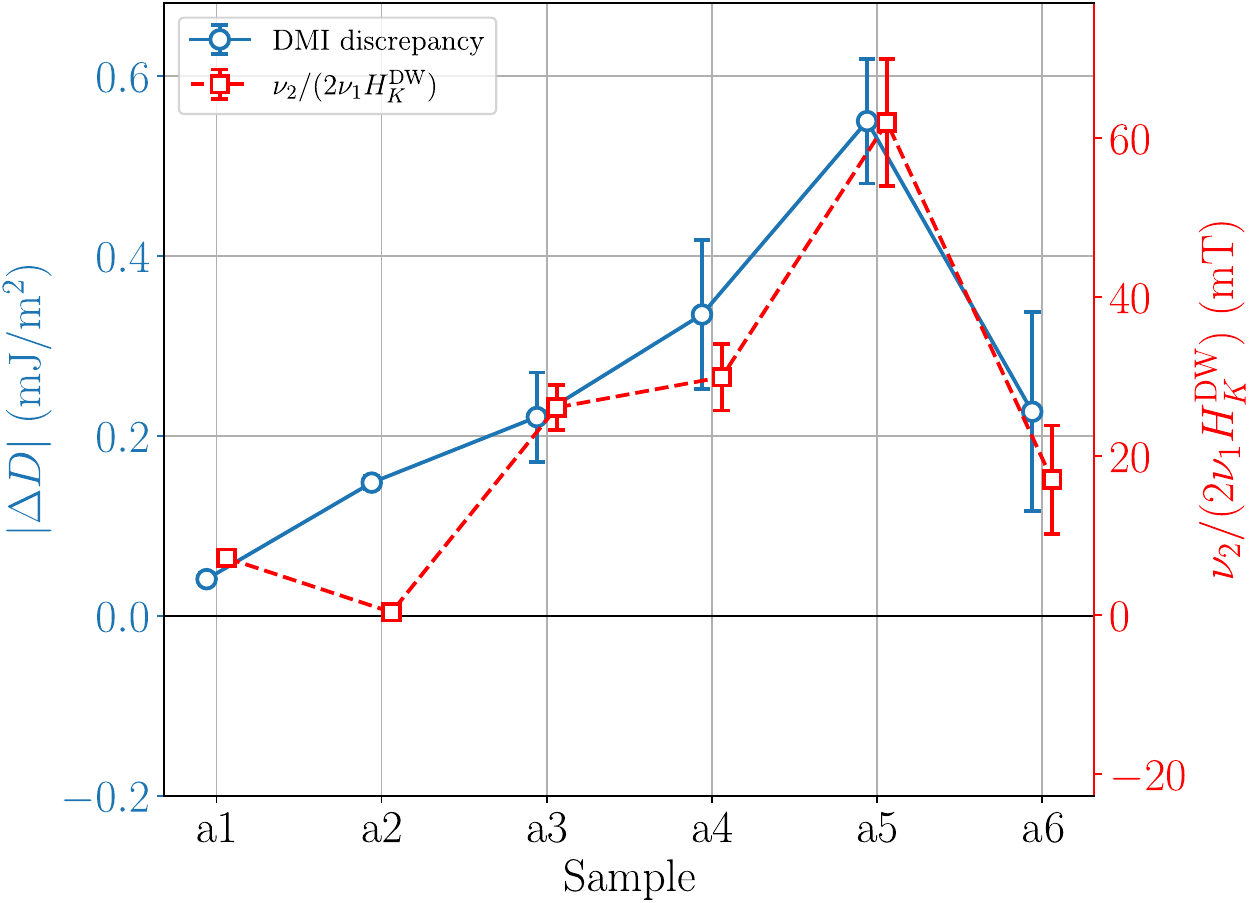}
    \caption{Comparison of the DMI discrepancy between BLS and conventional creep-regime measurements (left axis) with the fitted shift scale \(\nu_2/(2\nu_1H_K^{\mathrm{DW}})\) (right axis). Here, \(H_K^{\mathrm{DW}}\) denotes the crossover field between mixed Bloch–Néel and saturated Néel domain-wall configurations. Error bars represent propagated uncertainties; lines are guides to the eye.}
    \label{fig:dmi_discrepancy_dual_axis}
\end{figure}

A useful check for the model validity is provided by comparing the DMI values discrepancy and the predicted shift of Eq.(11) in the main text. As can be seen in Fig.~\ref{fig:dmi_discrepancy_dual_axis}, there is a clear correlation between the discrepancy of the DMI values $|\Delta D|$ (see Table.~\ref{tab:DMI_comparison}) and the factor $\nu_2/(2\nu_1 H_K^{DW})$ which quantifies the displacement of the DMI field from the velocity curve minimum.

\begin{table*}[htbp]
    \centering
    \caption{
    Fitting parameters obtained from the three-amplitude creep model.
    We report the applied bubble expansion fields $\mu_0 H_z$, the dimensionless amplitudes $\bar{\nu}_{1\dots3}$,
    the fitted crossover field $\mu_0H_{K,\mathrm{fit}}^{DW}$, and the fitted DMI effective
    field $\mu_0H_{\mathrm{DMI}}^{\mathrm{new}}$.
    The uncertainties in $\bar{\nu}_1$ and $\bar{\nu}_3$ include the
    propagated uncertainty in $\mu_0H_{K,\mathrm{fit}}^{DW}$, assuming the fitted quantities
    are statistically independent. We also include the effective DMI field measured in the flow regime $\mu_0H_{\mathrm{DMI}}^{\mathrm{flow}}$ (see \ref{subsec:DMI field from in-plane-field compensation}) for comparison.
    }
    \label{tab:three_amplitude_parameters}
    \begin{tabular}{@{}llcccccc@{}}
        \hline\hline
        Sample
         &$\mu_0 H_z$& $\bar{\nu}_1$
        & $\bar{\nu}_2$
        & $\bar{\nu}_3$
        & $\mu_0H_{K,\mathrm{fit}}^{DW}$
        & $\mu_0H_{\mathrm{DMI}}^{\mathrm{new}}$  
        & $\mu_0H_{\mathrm{DMI}}^{\mathrm{flow}}$ \\
         &(mT)& ($-$)
        & ($-$)
        & ($-$)
        & (mT)
        & (mT)  
        & (mT)  \\
        \hline

        $a1$
         &12.6& $1.0 \pm 0.2$& $0.94 \pm 0.05$& $0.78 \pm 0.08$& $15 \pm 1$& $- 2 \pm 1$&  $\approx 0$             \\

        $a2$
         &11.2& $7 \pm 2$& $0.14 \pm 0.09$& $4.1 \pm 0.5$
        & $40 \pm 4$
        & $4 \pm 2$ &$\approx 0$             \\

        $a3$
         &10.5& $0.7 \pm 0.1$& $0.69 \pm 0.03$& $0.83 \pm 0.06$
        & $55 \pm 4$
        & $-62 \pm 2$ &$-70\pm 7$        \\

        $a4$
         &12.4& $0.37 \pm 0.07$
        & $0.36 \pm 0.03$
        & $0.37 \pm 0.03$
        & $62 \pm 5$
        & $-69 \pm 5$  &$-88\pm 9$        \\

        $a5$
         &35& $1.3 \pm 0.2$& $2.36 \pm 0.08$
        & $0.71 \pm 0.04$
        & $70 \pm 3$& $-144 \pm 4$  &$-160\pm16$        \\

        a6
         &20& $0.7 \pm 0.2$& $0.22 \pm 0.08$
        & $1.5 \pm 0.2$& $113 \pm 13$
        & $88 \pm 7$  &n.m.                     \\

        \hline\hline
    \end{tabular}
\end{table*}

\subsection*{Smoothing procedure and bootstrap uncertainty}
\label{sec:bootstrap}
\noindent
The fitting procedure used to extract \(H_\mathrm{DMI}\) from the creep-regime velocity curves is based on the three-amplitude model derived above. In this model the logarithmic velocity is written as the sum of a symmetric and an antisymmetric contribution with respect to the shifted field coordinate
\begin{equation}
    X = H_x + H_\mathrm{DMI}.
\end{equation}
The ideal analytical form contains piecewise branches separated by the boundaries
\begin{equation}
    X = \pm H_K^{DW},
\end{equation}
which delimit the mixed Bloch-Néel region. While this piecewise form is useful for identifying the physical origin of the different contributions, the sharp transition at the boundaries can make numerical error estimation unstable. In particular, small changes of the fitting parameters can move experimental points across the boundary between branches, producing a non-smooth local optimization landscape. To avoid this purely numerical instability, the antisymmetric contribution was regularized by replacing the sharp transition between the central linear branch and the saturated outer branches with a smooth interpolation. The unsmoothed form is
\begin{equation}
    (\ln v)_A =
    \begin{cases}
        -\nu_2, & X < -H_K^{DW}, \\[4pt]
        \nu{}_2 X/H_K^{DW}, & |X| \leq H_K^{DW}, \\[4pt]
        \nu_2, & X > H_K^{DW}.
    \end{cases}
\end{equation}
In the numerical implementation, the three branches were blended using smooth switching functions of the form
\begin{equation}
    S_\pm(X)
    =
    \frac{1}{2}
    \left[
    1+\tanh\left(\frac{\pm X-H_K^{DW}}{w}\right)
    \right],
\end{equation}
where \(w\) is a fixed smoothing width. The smoothed antisymmetric contribution can then be written schematically as
\begin{equation}
    (\ln v)_A^\mathrm{smooth}
    =
    S_-(X)(-\nu_2)
    +
    \left[1-S_-(X)-S_+(X)\right]\nu_2\frac{X}{H_K^{DW}}
    +
    S_+(X)\nu_2 .
\end{equation}
The smoothing width was kept small compared with the field scale over which the velocity varies and was not treated as a free fitting parameter. Its role is only to regularize the branch boundaries for numerical minimization and uncertainty estimation.

Parameter uncertainties were estimated using a residual bootstrap. First, the model was fitted to the experimental data, yielding the best-fit parameter vector
\begin{equation}
    \hat{\theta}
    =
    \left(
    H_\mathrm{DMI},
    \nu_1,
    \nu_2,
    \nu_3,
    H_K^{DW}
    \right).
\end{equation}
The fitted values are
\begin{equation}
    \hat{y}_i = f(H_{x,i};\hat{\theta}),
\end{equation}
where \(y_i=\ln v_i\), and the residuals are
\begin{equation}
    r_i = y_i-\hat{y}_i .
\end{equation}
A synthetic bootstrap dataset was generated by sampling the residuals with replacement,
\begin{equation}
    y_i^\ast = \hat{y}_i + r_i^\ast ,
\end{equation}
where each \(r_i^\ast\) is randomly drawn from the original residual set. The model was then refitted to each synthetic dataset, producing a bootstrap parameter vector \(\theta^\ast\). Repeating this procedure \(N_\mathrm{boot}\) times gives an ensemble
\begin{equation}
    \left\{
    \theta_1^\ast,
    \theta_2^\ast,
    \ldots,
    \theta_{N_\mathrm{boot}}^\ast
    \right\}.
\end{equation}
The statistical uncertainty of each fitted parameter was estimated from the spread of this ensemble. For the DMI field, we report
\begin{equation}
    u({H_\mathrm{DMI}})
    =
    \mathrm{std}
    \left(
    H_{\mathrm{DMI},1}^{\ast},
    H_{\mathrm{DMI},2}^{\ast},
    \ldots,
    H_{\mathrm{DMI},N_\mathrm{boot}}^{\ast}
    \right).
\end{equation}
Since the bootstrap distribution was approximately symmetric, this standard deviation was used as the one-sigma statistical uncertainty. The quoted uncertainty should therefore be interpreted as the statistical fitting uncertainty within the adopted smoothed three-amplitude model.
\subsection{Uncertainty estimation of D}
The uncertainty in the $D$ value is then obtained through standard uncertainty propagation of the relevant physical quantities through the formula $D = \mu_0 M_s \Delta H_{\mathrm{DMI}}$, i.e. 
\begin{equation}
    u(D) = |D| \sqrt{\bigg[\frac{u(M_s)}{M_s}\bigg]^2 + \bigg[\frac{u(\Delta)}{\Delta}\bigg]^2 +  \bigg[\frac{u(H_{\mathrm{DMI}})}{H_{\mathrm{DMI}}}\bigg]^2 }.
\end{equation}
Given the rather complex ensemble of experimental data, we summarize here the source of all the relevant quantities for each sample. For a1 and a2, $M_s,u(M_s)$ and $\Delta, u(\Delta)$ are taken from Ref.\cite{magni_key_2022}. For a3, a4 and a5, $M_s,u(M_s)$ is taken from Ref.\cite{magni_key_2022}, while $\Delta$ and $u(\Delta)$ are taken from the comparison of flow measurements as discussed in Sec.\ref{sec:flow_characterization}. For sample a6, $M_s, u(M_S)$ and $K_{\mathrm{eff}},u(K_{\mathrm{eff}})$ are measured while the exchange stiffness $A , u(A)$  is taken from a literature survey on similar stacks \cite{devolder_exchange_2016,Zimmermann2019}. The DW width $\Delta$ and its standard uncertainty $u(\Delta)$ are obtained by propagating the uncertainty in $A$ and $K_{\mathrm{eff}}$ in $\Delta = \sqrt{A/K_{\mathrm{eff}}}$.  
\begin{equation}
    u(\Delta) = \frac{\Delta}{2} \sqrt{\bigg[\frac{u(A)}{A}\bigg]^2 + \frac{u(K_{\mathrm{eff}})}{K_{\mathrm{eff}}}\bigg]^2}.
\end{equation}

\begin{table*}[t]
    \centering
    \caption{DMI values obtained from BLS ($D_{\mathrm{BLS}}$), Walker-field ($D_{\mathrm{Walker}}$),
    conventional-creep \cite{Je2013} ($D_{\mathrm{creep}}^{\mathrm{old}}$) and new-creep analyses  ($D_{\mathrm{creep}}^{\mathrm{new}}$)}.
    \label{tab:DMI_comparison}
    \begin{tabular}{lcccc}
        \toprule
        Sample &
        $D_{\mathrm{BLS}}$~\cite{magni_key_2022} &
        $D_{\mathrm{Walker}}$ &
        $D_{\mathrm{creep}}^{\mathrm{old}}$&
        $D_{\mathrm{creep}}^{\mathrm{new}}$ \\
 & (mJ/m$^2$)& (mJ/m$^2$)& (mJ/m$^2$)&(mJ/m$^2$)\\
        \midrule
        a1     & $ 0.084 \pm 0.005$ & n.o.                 & $ 0.043 \pm 0.005$ & $ 0.060 \pm 0.008$ \\
        a2     & $ 0.170 \pm 0.006$ & n.o.                 & $ 0.022 \pm 0.004$ & $ 0.05 \pm 0.02$\\
        a3     & $-0.390 \pm 0.006$ & $-0.45 \pm 0.05$& $-0.24 \pm 0.06$& $-0.40 \pm 0.06$\\
        a4     & $-0.560 \pm 0.008$ & $-0.51 \pm 0.05$& $-0.3 \pm 0.1$& $-0.40 \pm 0.08$\\
        a5     & $-0.810 \pm 0.005$ & $-0.91 \pm 0.09$& $-0.45 \pm 0.06$& $-0.8 \pm 0.1$\\
        a6  & $ 0.8  \pm 0.1$& n.m.                 & $ 0.5  \pm 0.1 $& $ 0.7  \pm 0.2$\\
        \bottomrule
    \end{tabular}
\end{table*}

%